\documentclass{IEEEtran}
\usepackage{booktabs}
\usepackage{cite}
\usepackage{amssymb,amsfonts,amstext}% Lots of math symbols and enviro
\usepackage{amsmath}
\usepackage{graphicx} % For including graphics N.B. pdftex graphics dr
\usepackage{algorithm}
\usepackage{algorithmic}
\usepackage{multirow}
\usepackage[caption=false,font=footnotesize]{subfig}
\usepackage{stfloats}
\usepackage{url}
\usepackage{xcolor}
\usepackage{bbm}
\usepackage{psfrag}
\DeclareMathOperator*{\mini}{minimize}

\DeclareMathOperator{\sbto}{subject \text{ } to}
\DeclareMathOperator{\st}{s.t.}

\newtheorem{theorem}{Theorem}[section]

\newcommand{\qed}{\nobreak \ifvmode \relax \else
      \ifdim\lastskip<1.5em \hskip-\lastskip
      \hskip1.5em plus0em minus0.5em \fi \nobreak
      \vrule height0.75em width0.5em depth0.25em\fi}

\ifCLASSINFOpdf
\else
\fi
\usepackage{fixltx2e}
\usepackage{stfloats}

\begin{document}
\title{Energy Efficiency of Downlink Transmission Strategies for Cloud Radio Access Networks}

\author{\IEEEauthorblockN{Binbin Dai, \IEEEmembership{Student Member, IEEE} and Wei Yu, \IEEEmembership{Fellow, IEEE}}
%\IEEEauthorblockA{Department of Electrical and Computer Engineering\\
%         University of Toronto, Toronto, Ontario M5S 3G4, Canada  \\
%				Emails: \{bdai, weiyu\}@comm.utoronto.ca}
%\and
%\IEEEauthorblockN{Homer Simpson}
%\IEEEauthorblockA{Twentieth Century Fox\\
%\and
%\IEEEauthorblockN{James Kirk\\ and Montgomery Scott}
%\IEEEauthorblockA{Starfleet Academy\\
%San Francisco, California 96678-2391\\
%Fax: (888) 555--1212}

%\thanks{The materials in this paper have been presented in part at 
%the IEEE International Workshop on Signal Processing Advances in 
%Wireless Communications (SPAWC), Toronto, Canada, June 2014, \cite{binbin14}.
%This work was supported by Huawei Technologies Canada and by
%Natural Sciences and Engineering Research Council (NSERC) of Canada.
%The authors are with the Edward S. Rogers Sr. Department of Electrical
%and Computer Engineering, University of Toronto, Toronto, ON M5S 3G4, 
%Canada. (e-mails: bdai@ece.utoronto.ca, weiyu@comm.utoronto.ca).}

\thanks{Manuscript submitted on April 15, 2015; revised on September 15, 2015; accepted on December 11, 2015. This work was supported by Huawei Technologies, Canada, 
and Natural Sciences and Engineering Research Council (NSERC) of Canada. 
The authors are with The Edward S. Rogers Sr. Department of Electrical
and Computer Engineering, University of Toronto, Toronto, ON M5S 3G4, 
Canada (e-mails: \{bdai, weiyu\}@comm.utoronto.ca).}
}


%Telephone: (800) 555--1212, Fax: (888) 555--1212}


\maketitle

\begin{abstract}

This paper studies the energy efficiency of the cloud radio access network (C-RAN), specifically focusing on two fundamental and different downlink transmission strategies, 
namely the data-sharing strategy and the compression strategy. 
In the data-sharing strategy, the backhaul links connecting the central processor (CP) and the base-stations (BSs) are used to carry user messages -- each user's messages 
are sent to multiple BSs; the BSs locally form the beamforming vectors then cooperatively transmit the messages to the user. 
In the compression strategy, the user messages are precoded centrally at the CP, which forwards a compressed version of the analog beamformed signals to the BSs for cooperative transmission. 
This paper compares the energy efficiencies of 
the two strategies by formulating an optimization problem of minimizing the total network power consumption subject to user target rate constraints, 
where the total network power includes the BS transmission power, BS activation power, and load-dependent backhaul power. 
To tackle the discrete and nonconvex nature of the optimization problems, 
we utilize the techniques of reweighted $\ell_1$ minimization and successive convex approximation 
to devise provably convergent algorithms. 
Our main finding is that both the optimized data-sharing and compression strategies in C-RAN achieve much higher energy efficiency as compared to the non-optimized coordinated multi-point transmission, 
but their comparative effectiveness in energy saving depends on the user target rate. 
At low user target rate, data-sharing consumes less total power than compression, however, as the user target rate increases, 
the backhaul power consumption for data-sharing increases significantly leading to better energy efficiency of compression at the high user rate regime.

\end{abstract}


% \fi
\IEEEpeerreviewmaketitle

\section{Introduction}

\IEEEPARstart{U}{ltra-dense} deployment of small cells and cooperative communications are recognized as 
two promising technologies to meet the ever increasing demand of data traffic for future wireless networks \cite{Rost14}. 
However, both technologies come at the cost of increase in energy consumption 
because of the additional energy needed to support the increasing number of base-station (BS) sites and the substantially increased backhaul between the BSs for cooperation. 
The excessive energy consumption of wireless networks not only has an ecological impact in terms of carbon footprint but also has 
an economical impact on the operational cost to the mobile operators.  
Thus, the compelling call for improvement of \emph{spectrum efficiency} in the fifth-generation (5G) wireless network 
needs to be accompanied by a call for improvement of \emph{energy efficiency} to the same extent.

Cloud radio access network (C-RAN) is an emerging network architecture that shows significant promises in improving both the 
spectrum efficiency and the energy efficiency of current wireless networks \cite{CRAN}. 
In C-RAN, the BSs are connected to a central processor (CP) through backhaul links. 
The benefits of the C-RAN architecture in energy saving are several-fold. 
First, under the C-RAN architecture, most of the baseband signal processing in traditional BSs 
can be migrated to the cloud computing center 
so that the conventional high-cost high-power BSs can be replaced by low-cost low-power radio remote heads (RRHs). 
Second, the existence of CP also allows for the joint precoding of user messages for interference mitigation. 
With less interference generated, the transmit power at the BSs can therefore be reduced. 
Third, as on average (and especially during non-peak time) a significant portion of network resources can be idle \cite{Correia10}, 
the CP can perform joint resource allocation among the BSs to allocate resources on demand and put idle BSs into 
sleep mode for energy saving \cite{Frenger11}.

The above-mentioned benefits of C-RAN in energy saving are concerned with the BS side. However, the additional energy consumption due to 
the increased backhaul between the CP and the BSs also needs to be taken into account \cite{Tombaz11}. 
In this paper, we investigate the potential of C-RAN in improving energy efficiency of the 
communication aspect of the network by considering 
the energy consumption due to BS activation, transmission, and backhaul provisioning. 
The backhaul energy consumption depends on the backhaul rate, which further 
depends on the interface between the CP and the BSs. 
In this paper, we investigate two fundamental and different transmission strategies for the downlink C-RAN. 
In the \emph{data-sharing strategy}, the CP uses the backhaul links to share user messages to a cluster of cooperating BSs. 
The backhaul cost of the data-sharing strategy depends on the number of BSs that the user messages need to be delivered to:
larger cluster size leads to larger cooperative gain, but also higher backhaul rate. 
In an alternative strategy called the \emph{compression strategy}, the CP performs joint precoding of the user messages centrally then forwards a compressed version of the 
precoded signals to the BSs.
The backhaul cost of the compression strategy depends on the resolution of 
the compressed signals: higher-resolution leads to better beamformers, but also larger backhaul rate. 

This paper aims to quantify the energy saving of C-RAN while accounting for both the BS and the backhaul energy consumptions, and 
specifically to answer the question of between the data-sharing strategy and the compression strategy, which one is more energy efficient? 
The answer to this question is nontrivial as there are three factors that can lead to energy reduction: 
decrease in BS transmit power, turning-off of the BSs, and reduction in backhaul rate. 
These three factors are interrelated. For example, it may be beneficial to keep more BSs active and to allocate higher backhaul 
rate in order to allow for better cooperation among the BSs so that more interference can be mitigated. 
This leads to less transmit power consumption, but it can also lead to higher BS and backhaul power consumption. 
This paper intends to capture such interplay using an optimization framework. 
Specifically, we propose a joint design of the BS transmit power, BS activation and backhaul by minimizing 
network-wide power under given user rate constraints for both the data-sharing strategy and the compression strategy. 
The resulting optimization problems are nonconvex in nature and are highly nontrivial to solve globally. 
This paper approximates the problems using reweighted $\ell_1$ minimization technique \cite{Candes08} 
and successive convex approximation technique, and devises efficient algorithms with convergence guarantee. 
We identify operating regimes where one strategy is superior to the other, and show that overall optimized C-RAN transmission can lead to 
more energy efficient network operation than the non-optimized coordinated multi-point (CoMP) transmission.

\subsection{Related Work}

The potential of C-RAN in improving the performance for future wireless networks has 
attracted considerable attentions recently. 
In the uplink, \cite{Lei13} and \cite{Yuhan14} show that with the capability of jointly decoding user messages in the CP, 
the throughput of traditional cellular networks can be significantly improved. 
A similar conclusion also has been drawn in the downlink. 
In particular, \cite{Park13} proposes a joint design of beamforming and multivariate compression to 
maximize the weighted sum rate for the compression strategy, while 
\cite{hong12} and \cite{BinbinSparseBFJnal} consider joint beamforming and BS clustering design to maximize the 
weighted sum rate for data-sharing.

This paper focuses on the energy efficiency of C-RAN in the downlink.  
Several metrics have been proposed in the literature to measure the energy efficiency of a network. 
For example, the area power consumption metric (watts$/$unit area) is proposed in \cite{Richter09} to evaluate 
the energy efficiency of networks of different cell site densities. 
Another widely adopted measurement in energy efficiency is bits per joule metric, 
which has been studied in \cite{Sarkiss12} under a simplified 
single-user-two-BSs model from an information-theoretical point of view and 
also studied in \cite{Schober12, Peng14, Huq14} for 
orthogonal frequency division multiple access (OFDMA) based cooperative networks from the practical system design point of view.

In this paper, we formulate the problem of minimizing the total required power for downlink C-RAN in order to 
provide a given set of quality-of-service (QoS) targets for the scheduled users. 
Such an optimization problem can also be thought of as minimizing watts per bit (or maximizing the bits per watt) at given user service rates. 
In this domain, most of the previous works are restricted to the data-sharing strategy \cite{Shi13, Rui14, Han14}. 
Specifically, \cite{Shi13} proposes a joint BS selection and beamforming design algorithm to minimize the total power consumption in the 
downlink, while \cite{Rui14} generalizes to a joint downlink and uplink total power minimization problem. 
Both \cite{Shi13} and \cite{Rui14} take advantage of the fact that if a BS is not selected to serve any user at the current time slot, 
it can be put into low-power sleep mode for energy saving purpose.  
In contrast, \cite{Han14} assumes fixed BS association but exploits the delay tolerance of the users to improve the energy efficiency 
in CoMP transmission. 
In delay-tolerant applications, BSs can aggregate the user messages and transmit them with high rate during a short time frame while 
remain idle for the rest of the time slots under power-saving sleep mode. 
Fast deactivation/activation of hardware power-consuming components achieve significant energy reductions \cite{Frenger11, Kimmo13}.

In this paper, we adopt a similar energy saving perspective as in \cite{Shi13} but consider in addition the compression strategy. 
Compression strategy differs from data-sharing strategy in the way that the backhaul is utilized. 
We adopt the model proposed in \cite{Fehske10} to model the power consumption of backhaul links as a linear function of backhaul rates. 
This is in contrast to \cite{Shi13}, where the backhaul power is modeled as a step function 
with only two levels of power consumption depending on whether the backhaul link is active or not.

In addition to the backhaul power, we also consider the BS power consumption 
by adopting the model proposed in \cite{Auer11}, which 
approximates the power consumption of a BS as a piecewise linear function of transmit power. 
In such model, BS sleep mode corresponds to a constant but lower power consumption with zero transmit power. 
BS active mode corresponds to a higher constant power plus a nonzero transmit power. 
The overall framework of this paper is a joint optimization of BS transmit power, BS activation and backhaul rate for both the compression and the data-sharing strategies.

From the optimization perspective, the total power consumption for the data-sharing strategy 
involves a sum of weighted nonconvex $\ell_0$-norms, 
which is highly nontrivial to optimize globally. 
Instead, we adopt the reweighted $\ell_1$ technique \cite{Candes08} to approximate 
the nonconvex total power into a convex weighted sum of transmit power, 
where the weights are iteratively updated in a way to reduce not only the number of active BSs 
but also the backhaul rate. 
Such technique has also been applied to minimize the total backhaul rate in \cite{Zhao12}, 
and to optimize the tradeoff between the total transmit power and total backhaul rate in \cite{binbin13}. 
On a related note, the discrete $\ell_0$-norm can also be approximated using other tractable continuous functions such as 
Gaussian-like function in \cite{Zhuang14} and exponential function in \cite{Vu14}. 
It has been reported recently in \cite{ZhouTaoChen} that 
those approximation methods show similar effectiveness in inducing sparsity.

Further, the mathematical expression of the total power consumption for the compression strategy involves a difference of two logarithmic functions, 
which is also nonconvex. 
We propose to approximate the first logarithmic function using the successive convex approximation technique, which transforms the objective function into a convex form. 
The adopted reweighted $\ell_1$ minimization technique and successive convex approximation technique in this paper are related to 
the majorization-minimization (MM) algorithm \cite{Bharath11}, 
which deals with an optimization problem with nonconvex objective function by successively solving a sequence of optimization 
problems with approximate objective functions. 
This paper utilizes the known sufficient conditions of convergence for the MM algorithm in the literature \cite{Meisam13} 
to show the convergence of the proposed algorithms for both the data-sharing and the compression strategies.

%{\color{blue}
Finally, we mention that the data-sharing and compression strategies considered in this paper are not the only possibilities for the downlink of C-RAN. 
There is a potential to combine these two strategies by sending directly the messages of only the strong users to the BSs and compressing the rest \cite{Patil14}. 
Also, reverse compute-and-forward strategy that accounts for the lattice nature of the transmitted message is also possible
\cite{HongCaire13}. However, such strategy is difficult to optimize because of the need in choosing the 
right integer zero-forcing precoding coefficients at the CP so that the effective noise, caused by the non-integer penalty due to practical channels, at each user is minimized. %}

\subsection{Main Contributions}

This paper considers energy-efficient design of the data-sharing strategy and the compression strategy for 
downlink C-RAN by formulating a problem of minimizing the total network power consumption subject to user rate constraints. 
The first contribution of this paper is the modeling of both the BS power and the 
backhaul power consumption in the network. 
The BS power consumption model includes a low-power sleep mode, while the backhaul power consumption 
is modeled as a linear function of backhaul traffic rate. 

For the data-sharing strategy, we propose a novel application of reweighted $\ell_1$ minimization technique to approximate the 
nonconvex BS activation power and backhaul power. 
Such approximation technique reduces the nonconvex optimization problem to a conventional convex 
transmit power minimization problem, 
which can be solved efficiently using the uplink-downlink duality approach or through transformation as 
second-order cone programming (SOCP). Moreover, we adopt a reweighting function that enables us to 
connect the reweighted $\ell_1$ minimization technique with the MM algorithm. 
This connection allows us to prove the convergence behavior of the proposed algorithm for the data-sharing strategy.  

For the compression strategy, in addition to the reweighted $\ell_1$ approximation to the BS activation power as in the data-sharing strategy, 
we propose a successive convex approximation to the backhaul power, which is in a nonconvex form as 
a difference of two logarithmic functions. 
The proposed successive convex approximation technique and the reweighted $\ell_1$ approximation technique can be 
combined together. 
The combined algorithm falls into the class of the MM algorithms and has convergence guarantee.

Through simulations, we show that optimized data-sharing and compression strategies in C-RAN can bring much improved energy efficiency as compared to the non-optimized CoMP transmission.
However, the comparative energy saving of data-sharing 
versus compression depends on the user target rates. 
The energy efficiency of the data-sharing strategy is superior to that of the compression strategy in the low-rate regime. However, the  backhaul power consumption of the data-sharing strategy 
increases significantly with the user rate. Thus, in high user rate regime, the compression strategy may be preferred from an energy saving perspective.

\subsection{Paper Organization and Notations}

The remainder of this paper is organized as follows. 
Section~\ref{sec:SystemModel} introduces the system model and power consumption model considered throughout this paper. 
Section~\ref{sec:data_sharing} considers the total power minimization under the data-sharing transmission strategy, while 
Section~\ref{sec:compression} considers the compression strategy. 
Simulation results are presented in Section~\ref{sec:simulations} and conclusions are drawn in Section~\ref{sec:conclusion}.

Throughout this paper, lower-case letters (e.g. $x$) and lower-case bold letters (e.g. $\mathbf{x}$) denote scalars and column vectors 
respectively. 
We use $\mathbb{C}$ to represent complex domain. 
The transpose, conjugate transpose and $\ell_p$-norm of a vector are denoted as $(\cdot)^{T}$, $(\cdot)^H$ and $\Vert\cdot\Vert_p$ respectively. 
The expectation of a random variable is denoted as $\mathsf{E} \left [ \cdot \right]$. 
Calligraphy letters are used to denote sets, while $|\cdot|$ stands for either the size of a set or the absolute value of a  scalar, depending on the context.

%This demo file is intended to serve as a ``starter file''
%for IEEE conference papers produced under \LaTeX\ using
%IEEEtran.cls version 1.7 and later.
%% You must have at least 2 lines in the paragraph with the drop letter
%% (should never be an issue)
%I wish you the best of success.
%\hfill January 11, 2007
%
%\subsection{Subsection Heading Here}
%Subsection text here.
%
%
%\subsubsection{Subsubsection Heading Here}
%Subsubsection text here.

% An example of a floating figure using the graphicx package.
% Note that \label must occur AFTER (or within) \caption.
% For figures, \caption should occur after the \includegraphics.
% Note that IEEEtran v1.7 and later has special internal code that
% is designed to preserve the operation of \label within \caption
% even when the captionsoff option is in effect. However, because
% of issues like this, it may be the safest practice to put all your
% \label just after \caption rather than within \caption{}.
%
% Reminder: the "draftcls" or "draftclsnofoot", not "draft", class
% option should be used if it is desired that the figures are to be
% displayed while in draft mode.
%
%\begin{figure}[!t]
%\centering
%\includegraphics[width=2.5in]{myfigure}
% where an .eps filename suffix will be assumed under latex, 
% and a .pdf suffix will be assumed for pdflatex; or what has been declared
% via \DeclareGraphicsExtensions.
%\caption{Simulation Results}
%\label{fig_sim}
%\end{figure}

% Note that IEEE typically puts floats only at the top, even when this
% results in a large percentage of a column being occupied by floats.

\section{System and Power Consumption Model}\label{sec:SystemModel}

In this section, we describe the overall system model and power consumption model for the downlink C-RAN 
considered throughout this paper. 

%
%\begin{figure}[t]
  %\centering
  %\includegraphics[width= 0.45\textwidth]{figs/perBS.eps}
%\caption{Downlink C-RAN with per-BS backhaul capacity limits, where each user is cooperatively served by a user-centric and potentially
%overlapping subset of BSs.}
%\label{fig:SystemModel}
%\end{figure}

\subsection{System Model}

Consider a downlink C-RAN with $L$ BSs serving $K$ users. 
All the BSs are connected to a CP via backhaul\footnote{This paper refers the link between CP and BSs as \emph{backhaul}, which 
is appropriate if the data-sharing strategy is used. 
However, in a C-RAN architecture implementing the compression strategy where the BSs are simply RRHs, 
the connection between RRH and CP can be referred to more appropriately as \emph{fronthaul}.} links and each user  receives a single independent data stream from the BSs. 
All the user messages are assumed to be available at the CP and are jointly processed before being forwarded to the BSs through the backhaul links. %{\color{blue}
We assume that the CP has access to global channel state information (CSI) but point out that such assumption can be relaxed so that 
only the CSI from the neighboring BSs of each user is needed in the CP. %}

To simplify notations and ease analysis, we assume that the BSs and the users are equipped with a single antenna each, although the 
proposed algorithms in this paper can be easily generalized to the case of multi-antenna BSs as discussed later in the paper.  
Let $x_l \in \mathbb{C}$ denote the transmit signal at BS $l$, we can write the received signal $y_k \in \mathbb{C}$ at user $k$ as 
\begin{equation}\label{eq:yk_general}
y_k = \mathbf{h}_k^{H} \mathbf{x} + n_k, \quad k \in  \mathcal{K} = \left\{1, 2, \cdots, K\right\}
\end{equation}
where $\mathbf{x} = \left[x_1, x_2, \cdots, x_L\right]^{T}$ is the vector of transmit signals across all the $L$ BSs and 
$\mathbf{h}_k \in \mathbb{C}^{L \times 1}$ is the vector of channel gains from all the $L$ BSs to user $k$. 
The received noise $n_k$ is modeled as a complex Gaussian random variable with zero mean and variance $\sigma^2$. 
Each user decodes its own message $s_k \in \mathbb{C}$ from the received signal $y_k$.

In this paper, we investigate two fundamental but different transmission strategies, 
the data-sharing strategy and the compression strategy, for the downlink C-RAN for delivering the message $s_k$ to user $k$ 
via the transmit signal $\mathbf{x}$ from the BSs. 
In particular, we compare the potential of these two strategies in improving the energy efficiency. 
Before discussing the details of the two strategies, we first describe the power consumption model adopted in this paper.

\subsection{Power Consumption Model}

Traditional cellular network transmission strategy design typically only considers transmit power at each BS, which is written as 
\begin{equation}\label{eq:TxPower}
P_{l, tx} = \mathsf{E} \left[ \vert x_l \vert^{2} \right] \leq P_l, \quad l \in \mathcal{L} = \left\{1, 2, \cdots, L\right\}, 
\end{equation}
where $P_l$ is the transmit power budget available at BS $l$. 
However, a full characterization of power consumption at a BS should also consider the efficiency of the power amplifier and 
other power-consuming components such as baseband unit, cooling system, etc. 
In addition, the power consumption of backhaul links connecting the BSs to the CP also needs to be taken into account for the specific 
C-RAN architecture considered in this paper. 
In the following, we describe the power consumption model adopted in this paper for the BSs and the backhaul links respectively.

\subsubsection{Base-Station Power Consumption}

The characteristic of power-consuming components in a BS depends on the BS design. 
We adopt the following unified power consumption model proposed in \cite{Auer11}, which is applicable for different types of BSs. 
This model approximates the BS power consumption as a piecewise linear function of the transmit power $P_{l, tx}$: 
\begin{equation} \label{eq:bs_power}
P_{l}^{BS} = \left \{
   \begin{array}{ll}
	 \eta_l P_{l, tx} + P_{l, active}, & \text{if} ~ 0 < P_{l, tx} \leq P_l \\
	 P_{l, sleep}, & \text{if} ~ P_{l, tx} = 0
	 \end{array}
  \right. \hspace{-3mm}, ~ l \in \mathcal{L}
\end{equation}
where $\eta_l > 0$ is a constant reflecting the power amplifier efficiency, feeder loss and other loss factors due to power supply and cooling for BS $l$, 
$P_{l, tx}$ is the transmit power defined in \eqref{eq:TxPower} and $P_{l, active}$ is the minimum power required to support BS $l$ with 
non-zero transmit power. If BS $l$ has nothing to transmit, it can be put into sleep mode with low power consumption $P_{l, sleep}$. 
Typically, $P_{l, sleep} < P_{l, active}$ so that it is beneficial to turn BSs into sleep mode, whenever possible, 
for energy-saving purpose.

\subsubsection{Backhaul Power Consumption}

In C-RAN, the BSs are connected to the CP with the backhaul links. 
The power consumption due to backhaul links varies with different backhaul technologies. 
In this paper, we model the backhaul as a set of communication channels, each with 
capacity $C_l$ and power dissipation $P_{l,max}^{BH}$, and write the backhaul power consumption as 
\begin{equation}\label{eq:bkhaul_power}
P_l^{BH} = \frac{R_l^{BH}}{C_l} P_{l,max}^{BH} = \rho_l R_l^{BH}, \quad l \in \mathcal{L}
\end{equation}
where $\rho_l = P_{l,max}^{BH} / C_l$ is a constant scaling factor and $R_l^{BH}$ is the backhaul traffic between BS $l$ and the CP.
This model has been used in \cite{Fehske10} for microwave backhaul links and can also be generalized to other 
backhaul technologies, such as passive optical network, fiber-based Ethernet, etc., as mentioned in \cite{wu2012green}. 
Note that \cite{Shi13} also considers the sleep mode capability for backhaul links. We point out that 
such consideration can be unified with 
$P_{l, active}$ and $P_{l, sleep}$ in the BS power consumption model \eqref{eq:bs_power}.

%In this paper, we adopt the model used in \cite{Fehske10}, which models the backhaul as a collection of microwave links of 
%capacity $C_l$ and power dissipation $P_l^{MW}$. The backhaul power is written as 
%\begin{equation}\label{eq:bkhaul_power}
%P_l^{BH} = \frac{R_l^{BH}}{C_l} P_l^{MW} = \rho_l R_l^{BH}, \quad l \in \mathcal{L}
%\end{equation}
%where $\rho_l = P_l^{MW} / C_l$ is a constant scaling factor and $R_l^{BH}$ is the backhaul traffic between BS $l$ and the CP. 
%This power consumption model can also be generalized to other backhaul technologies \cite{wu2012green}. 
%Note that \cite{Shi13} also considers the sleep mode capability for backhaul links. We point out that 
%such consideration can be unified with 
%$P_{l, active}$ and $P_{l, sleep}$ in the BS power consumption model \eqref{eq:bs_power}. 

\subsubsection{Total Power Consumption}

Based on the above BS power consumption model \eqref{eq:bs_power} and backhaul power consumption model \eqref{eq:bkhaul_power}, 
we can write the total power consumption $P_{total}$ for C-RAN as
\begin{align}\label{eq:total_power}
P_{total} 
& = \sum_{l \in \mathcal{L}} \left( P_{l}^{BS} + P_l^{BH} \right)\nonumber \\ 
& = \sum_{l \in \mathcal{L}} \bigg( \eta_l P_{l, tx} + \mathbbm{1}_{\left\{ P_{l, tx} \right\}} \left( P_{l, active} -  P_{l, sleep}  \right) \nonumber \\ 
& \hspace{4.5cm} + P_{l, sleep} + \rho_l R_l^{BH} \bigg) \nonumber \\
& = \sum_{l \in \mathcal{L}} \bigg( \eta_l P_{l, tx} + \mathbbm{1}_{\left\{ P_{l, tx} \right\}} P_{l, \Delta} + \rho_l R_l^{BH} \bigg) \nonumber \\ 
& \hspace{4.5cm} + \underbrace{\sum_{l \in \mathcal{L}} P_{l, sleep}}_{\text{constant}}
\end{align}
where $\mathbbm{1}_{\left\{ \cdot \right\}}$ is the indicator function defined as 
\begin{equation}\label{eq:indicator}
 \mathbbm{1}_{\left\{x\right\}}  = \left \{
   \begin{array}{l}
	 1, \quad \text{if} ~ x > 0 \\
	 0, \quad \text{otherwise} 
	 \end{array}
  \right. , 
\end{equation}
and $P_{l, \Delta} = P_{l, active} -  P_{l, sleep}$ is the difference between the minimum active BS power consumption and the 
sleep mode BS power consumption. 

As we can see from \eqref{eq:total_power}, there are three possibilities in improving the energy efficiency of C-RAN: 
reducing the transmit power, putting BSs into sleep mode, and decreasing the backhaul traffic. 
However, these three aspects cannot be realized simultaneously: deactivating more BSs means reduced capability for interference 
mitigation among the active BSs, which leads to higher transmit power in order to maintain the QoS for the users; 
higher backhaul rate can allow for more user information being shared among the BSs so that the BSs can better 
cooperate to mitigate interference, thus less transmit power may be needed. 
A joint design is necessary in order to balance the roles of transmit power, BS activation and backhaul traffic rate in achieving 
energy efficiency. 
In the following, we describe the general problem formulation considered in this paper for such joint design used in both the 
data-sharing and the compression strategies. 
%In this paper, we propose such joint design for both the 
%data-sharing strategy and the compression strategy under the framework of minimizing the total network power consumption 
%in \eqref{eq:total_power} subject to user target rate constraints. 

%{\color{blue}

\subsection{Energy Efficiency Maximization}

This paper aims to understand the energy efficiency for downlink C-RAN, which can be defined as 
the ratio of the achievable sum rate and the sum power consumption, i.e. $\frac{\sum_{k} R_k}{P_{total}}$ 
where $R_k$ is the data rate for user $k$ determined by the specific transmission strategy 
and $P_{total}$ is the total consumed power defined in \eqref{eq:total_power}. 
Towards this end, this paper takes the similar approach as in \cite{Han14} to fix the service rates of scheduled users 
and consider the minimization of total power consumption: 
\begin{align} \label{prob:power_min}
\mini & \quad  P_{total} \\
 \sbto & \quad R_k \geq r_k, \quad \forall k \in \mathcal{K} \nonumber \\
& \quad \mathsf{E} \left[ \vert x_l \vert^{2} \right] \leq P_l, \quad \forall l \in \mathcal{L} \nonumber 
\end{align}
where $r_k$ is the fixed target rate for user $k$. 
The solution to the above problem gives us the energy efficiency $\frac{\sum_{k} R_k}{P_{total}}$ of the system 
at the operating point $\left(r_1, r_2, \cdots, r_K \right)$. 
To maximize energy efficiency, we need to further search over all operating points. 
For the rest of the paper, we study and compare the minimum required total power for different transmission strategies under the same operating point $\left(r_1, r_2, \cdots, r_K \right)$ in the downlink of C-RAN. 
Note that problem \eqref{prob:power_min} implicitly assumes fixed user scheduling. 
There also exists a possibility of doing joint user scheduling and power minimization 
by considering a problem of minimizing the total power consumption across \emph{multiple time slots} subject to a minimum target 
for each user's \emph{average} rate. Such problem is considerably more complicated. 

%}

\subsection{Data-Sharing versus Compression}

Data-sharing and compression are two 
fundamentally different transmission strategies for the downlink of C-RAN for delivering data to the users. 
These two strategies correspond to alternative functional splits in C-RAN.
In the data-sharing transmission strategy, the CP routes each scheduled user's intended message to a cluster of BSs 
through the backhaul links; the cluster of BSs then cooperatively serve that user through joint beamforming. 
In contrast, in the compression strategy, the precoding operation is implemented centrally at the CP, which then 
forwards a compressed version of the analog beamformed signal to the BSs through the backhaul/fronthaul links. 
The BSs then simply transmit the compressed beamforming signals to the users \cite{Park13, Patil14, PratikEUSIPCO}. 
%For both strategies, most of the signal processing tasks are carried out in the CP with cloud computing capabilities rather than in 
%the traditional BSs, which can therefore be replaced by the low-power RRHs. 
%As a consequence, however, the CP requires additional power for implementing the excessive signal processing for the BSs. 
%We only focus on the power consumption due to the BSs and the backhaul links in this paper and refer to \cite{Chen14} for potential 
%interests in power consumption of the CP. 

The data-sharing strategy differs from the compression strategy in backhaul utilization.  
In data-sharing, the backhaul rate is a function of the user message rate and the BS cluster size, 
while in the compression strategy the backhaul cost is determined by the compression resolution. 
Intuitively, as the user target rate and BS cluster size increase, the backhaul rate for the data-sharing strategy 
would increase significantly, leading to high energy consumption. 
However, in the low user rate regime where the BS cluster size is small, data-sharing can be more efficient than compression as the latter suffers from quantization noise. 
Therefore, there exists a tradeoff between data-sharing and compression in terms of backhaul rate 
and energy efficiency at different user target rate operating points. 
In the following two sections, we describe in details the data-sharing strategy and the compression strategy, and propose 
corresponding algorithms to find the minimum required total power for each strategy.

Throughout this paper, we primarily account for the energy consumption due to 
\emph{communications} in either the backhaul or the transmission front-end at the BSs, 
rather than the energy consumption due to \emph{computing}. 
There is significant additional energy saving due to migrating signal processing from 
the BSs to the cloud computer center in the C-RAN architecture. We refer the readers to \cite{Chen14}.

\section{Data-Sharing Strategy}\label{sec:data_sharing}

\begin{figure}[!t]
  \centering
	\psfrag{p}[tc][Bc][1]{Central Processor}
	\psfrag{q}[tc][Bc][1]{$s_1, s_2$}
	\psfrag{a}[tc][cc][1]{$s_1$}
	\psfrag{b}[tc][cc][1]{$s_2$}
	\psfrag{i}[tl][tl][0.7]{$R_1^{BH} = R_1$}
	\psfrag{j}[tl][tl][0.7]{$R_2^{BH} = $}
	\psfrag{w}[tl][tl][0.7]{$R_1 + R_2$}
	\psfrag{k}[tl][tl][0.7]{$R_3^{BH} = R_2$}
	\psfrag{x}[tc][cl][0.8]{$x_1 = w_{11}s_1$}
	\psfrag{y}[tc][cl][0.8]{$x_2 = w_{21}s_1$}
	\psfrag{h}[tc][tl][0.8]{$+ w_{22}s_2$}
	\psfrag{z}[tc][cl][0.8]{$x_3 = w_{32}s_2$}
	\psfrag{m}[tc][tc][0.9]{BS $1$}
	\psfrag{e}[tc][tc][0.9]{BS $2$}
	\psfrag{g}[tc][tc][0.9]{BS $3$}
	\psfrag{n}[tc][tc][0.9]{User $1$}
	\psfrag{f}[tc][tc][0.9]{User $2$}
	
  \includegraphics[width= 0.45\textwidth]{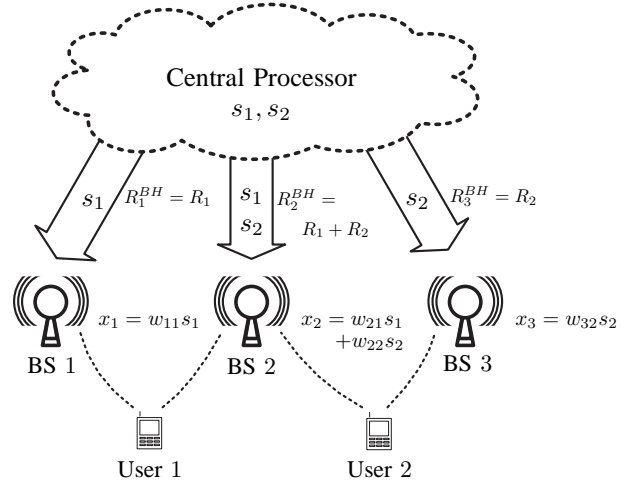}
\caption{Downlink C-RAN with data-sharing transmission strategy. In this illustrative example, the CP transmits 
user $1$'s message $s_1$ to BSs $1$ and $2$, and user $2$'s message $s_2$ to BSs $2$ and $3$. 
The BS cluster $(1, 2)$ then cooperatively serves user $1$ and the BS cluster $(2, 3)$ cooperatively serves user $2$ through joint beamforming.}
\label{fig:DataSharingModel} 
\end{figure}

In this section, we study the minimum total power required for the data-sharing strategy 
in order to support the given scheduled users at guaranteed service rates.

\subsection{Problem Formulation}

Consider the data-sharing transmission strategy for the downlink of C-RAN as illustrated 
in Fig.~\ref{fig:DataSharingModel}, where the each user's message is shared among a cluster of serving BSs. 
Let $w_{lk} \in \mathbb{C}$ be the beamforming coefficient for BS $l$ to serve user $k$. 
If BS $l$ is not part of user $k$'s serving cluster, $w_{lk}$ is set to be zero. 
The transmit signal $x_l$ at BS $l$ can be written as $x_l = \sum_{k \in \mathcal{K}} w_{lk} s_k$. 
We model the user messages $s_k$'s as independent and identically distributed complex Gaussian 
random variables with zero mean and unit variance. The transmit power $P_{l, tx}$ formulated in \eqref{eq:TxPower} can be written as 
\begin{equation}\label{eq:Tx_power_data_sharing}
P_{l, tx} = \sum_{k \in \mathcal{K}} \left\vert w_{lk} \right\vert^{2}, \quad l \in \mathcal{L}. 
\end{equation}

Substituting $x_l = \sum_{k \in \mathcal{K}} w_{lk} s_k$ into \eqref{eq:yk_general}, the received signal $y_k$ at user $k$ is 
\begin{equation}\label{eq:y_k_data_sharing}
y_k = \mathbf{h}_k^{H} \mathbf{w}_k s_k + \sum_{j \neq k}\mathbf{h}_k^{H} \mathbf{w}_j s_j + n_k, \quad k \in \mathcal{K}, 
\end{equation}
where $\mathbf{w}_k = \left[w_{1k}, w_{2k}, \cdots, w_{Lk}  \right]^{T}$ is the network beamformer for user $k$. 
Based on \eqref{eq:y_k_data_sharing}, the received signal-to-interference-plus-noise ratio (SINR) at user $k$ can be expressed as 
\begin{equation}
		\text{SINR}_k =
\frac{\left\vert\mathbf{h}_k^{H}\mathbf{w}_k \right\vert^2}{\sum_{j \neq
k}\left\vert\mathbf{h}_k^{H}\mathbf{w}_j\right\vert^2 + \sigma^2}, \quad k \in \mathcal{K}
\end{equation}
and the achievable rate for user $k$ is then
\begin{equation}\label{eq:R_k_data_sharing}
R_k = \log_2\left(1+\frac{\text{SINR}_k}{\Gamma_m}\right), \quad k \in \mathcal{K},  
\end{equation}
where $\Gamma_m$ stands for the signal-to-noise ratio (SNR) gap due to practical modulation scheme.

For the data-sharing strategy, if user $k$ is served by BS $l$, then the CP needs to send 
user $k$'s message $s_k$, along with the beamforming coefficient $w_{lk}$, to BS $l$ through the backhaul link. 
In this paper, we assume that the channels are slow varying and ignore the backhaul required for sharing CSI and beamformers, and only 
consider the backhaul capacity consumption due to data-sharing. 
Hence, the backhaul rate for BS $l$, $R_{l}^{BH}$, is the accumulated data rates of those users served by BS $l$, which 
can be formulated as 
\begin{equation} \label{eq:bkhaul_data_sharing}
R_{l}^{BH} = \sum_{k \in \mathcal{K}} \mathbbm{1}_{\left\{ \left\vert w_{lk} \right\vert^{2} \right\}} R_k, \quad l \in \mathcal{L}, 
\end{equation}
where $\mathbbm{1}_{\left\{ \left\vert w_{lk} \right\vert^{2} \right\}}$ is the indicator function defined in \eqref{eq:indicator} and 
indicates whether or not BS $l$ serves user $k$.

Substituting \eqref{eq:Tx_power_data_sharing} and \eqref{eq:bkhaul_data_sharing} into \eqref{eq:total_power}, 
the total power minimization problem \eqref{prob:power_min} can be formulated for the data-sharing strategy as
\begin{subequations} \label{prob1:data_sharing}
\begin{align} 
\displaystyle \mini_{\left\{w_{lk} \right\}} & \quad \sum_{l\in\mathcal{L}} 
\Bigg( \eta_l \sum_{k \in \mathcal{K}} \left\vert w_{lk} \right\vert^{2}  + 
\mathbbm{1}_{\left\{ \sum_{k \in \mathcal{K}} \left\vert w_{lk} \right\vert^{2} \right\}} P_{l, \Delta}   \nonumber \\ 
& \hspace{2.5cm} + \rho_l \sum_{k\in \mathcal{K}} \mathbbm{1}_{\left\{ \left\vert w_{lk} \right\vert^{2} \right\}} R_k \Bigg) 
\label{obj_data_sharing1} \\
 \sbto & \quad R_k = \log_2\left(1+\frac{\text{SINR}_k}{\Gamma_m}\right) \geq r_k, ~ k \in \mathcal{K} \label{rate_const} \\
  &\quad \sum_{k \in \mathcal{K}} \left\vert w_{lk} \right\vert^{2} \leq P_l, \quad l \in \mathcal{L}.
\end{align}
\end{subequations}
Note that the $\sum_{l \in \mathcal{L}} P_{l, sleep}$ term in \eqref{eq:total_power} is a constant and has been 
dropped in the objective function \eqref{obj_data_sharing1}.
It is easy to see that the minimum rate constraint \eqref{rate_const} is met with equality at the optimal point. 
Hence, problem \eqref{prob1:data_sharing} can be equivalently formulated as
\begin{subequations} \label{prob:data_sharing}
\begin{align} 
\displaystyle \mini_{\left\{w_{lk} \right\}} & \quad \sum_{l\in\mathcal{L}} 
\Bigg( \eta_l \sum_{k \in \mathcal{K}} \left\vert w_{lk} \right\vert^{2}  + 
\mathbbm{1}_{\left\{ \sum_{k \in \mathcal{K}} \left\vert w_{lk} \right\vert^{2} \right\}} P_{l, \Delta}  \nonumber \\ 
& \hspace{2.5cm} + \rho_l \sum_{k\in \mathcal{K}} \mathbbm{1}_{\left\{ \left\vert w_{lk} \right\vert^{2} \right\}} r_k \Bigg) 
\label{obj_data_sharing} \\
 \sbto  & \quad \text{SINR}_k \geq \gamma_k, \quad k \in \mathcal{K} \label{sinr_const} \\
&  \quad \sum_{k \in \mathcal{K}} \left\vert w_{lk} \right\vert^{2} \leq P_l, \quad l \in \mathcal{L} \label{Per_BS_Power}
\end{align}
\end{subequations}
where the variable $R_k$ in \eqref{obj_data_sharing1} is replaced by the target rate $r_k$ in \eqref{obj_data_sharing} and 
$\gamma_k = \Gamma_m \left(2^{r_k} - 1\right)$ in \eqref{sinr_const}. %{\color{blue}
The new SINR constraint \eqref{sinr_const} is also met with equality at the optimality. However, we keep \eqref{sinr_const} as an 
inequality constraint, so that it can be reformulated as a convex second-order cone (SOC) constraint \cite{wiesel2006}.
Note that problem \eqref{prob:data_sharing} is equivalent to problem \eqref{prob1:data_sharing} in the sense that they have the same 
optimal solutions and the same feasibility region. %}

Note that the above optimization is over the beamforming coefficients and also implicitly over the BS cluster for each user. 
The overall optimization problem \eqref{prob:data_sharing} aims to choose the optimal cluster of serving BSs for each scheduled user for 
minimizing the total power consumption while satisfying the user QoS constraints. 
Due to the indicator functions in the objective function \eqref{obj_data_sharing}, problem \eqref{prob:data_sharing} is nonconvex  
(discrete), 
so finding its global optimum solution is challenging. 
In the following, we propose to approximate the nonconvex indicator function using reweighted convex $\ell_1$-norm and show that 
with a particular reweighting function the proposed algorithm always converges\footnote{In fact, it converges to the stationary point solution of an approximation to problem \eqref{prob:data_sharing}.}.

\subsection{Proposed Algorithm}

We make an observation that the indicator function defined in \eqref{eq:indicator} is equivalent to the $\ell_0$-norm of a scalar. 
The $\ell_0$-norm of a vector is defined as the number of nonzero entries in the vector, so it reduces to the indicator function in 
the scalar case. In compressive sensing literature \cite{Candes08}, nonconvex $\ell_0$-norm minimization problem can be approximated 
as convex reweighted $\ell_1$ minimization problem. We take advantage of this technique and propose to approximate the 
indicator functions in the objective function \eqref{obj_data_sharing} as
\begin{equation}\label{wgt_l}
\mathbbm{1}_{\left\{ \sum_{k \in \mathcal{K}} \left\vert w_{lk} \right\vert^{2} \right\}}
= \left\Vert  \sum_{k \in \mathcal{K}} \left\vert w_{lk} \right\vert^{2} \right\Vert_0 
\approx \mu_l \sum_{k \in \mathcal{K}} \left\vert w_{lk} \right\vert^{2} 
\end{equation}
\begin{equation}\label{wgt_lk}
\mathbbm{1}_{\left\{ \left\vert w_{lk} \right\vert^{2} \right\}} 
= \left\Vert  \left\vert w_{lk} \right\vert^{2} \right\Vert_0
\approx \nu_{lk} \left\vert w_{lk} \right\vert^{2} 
\end{equation}
with weights $\mu_l$ and $\nu_{lk}$ iteratively updated according to 
\begin{align}\label{wgt_update_data}
 \mu_l = f\left( \sum_{k \in \mathcal{K}} \left\vert w_{lk} \right\vert^{2}, \tau_1 \right) 
& = \frac{c_1}{\sum_{k \in \mathcal{K}} \left\vert w_{lk} \right\vert^{2} + \tau_1} ~ , \nonumber \\ 
 \nu_{lk} = f\left(\left\vert w_{lk} \right\vert^{2}, \tau_2\right) 
& = \frac{c_2}{\left\vert w_{lk} \right\vert^{2} + \tau_2} %l \in \mathcal{L}, k \in \mathcal{K}
\end{align}
where $\left\{ w_{lk} \right\}$ is the beamformer from the previous iteration, $\tau_1 > 0$ and $\tau_2 > 0$ are some 
constant regularization factors, and $c_1, c_2$ are constants.  
%Here, $f\left(x, \tau\right)$ with parameter $\tau>0$ is a reweighting function to be specified later and is typically chosen 
%to be inversely proportional to $x$. 

%where $\mu_l$, $\nu_{lk}$ in \eqref{wgt_l}, \eqref{wgt_lk} are constant weights and updated iteratively according to 
%a reweighting function $f(x, \tau)$
%
%\begin{equation}\label{wgt_update_data}
%\mu_l = \frac{1}{\sum_{k \in \mathcal{K}} \left\vert w_{lk} \right\vert^{2} + \tau_1} ~, \quad 
%\nu_{lk} = \frac{1}{\left\vert w_{lk} \right\vert^{2} + \tau_2} %l \in \mathcal{L}, k \in \mathcal{K}
%\end{equation}
%with some small constant regularization factors $\tau_1 > 0$ and $\tau_2 > 0$ respectively. 

Note that in the above iterative updates of $\mu_l$ and $\nu_{lk}$, the BSs with small transmit power,  
$\sum_{k \in \mathcal{K}} \left\vert w_{lk} \right\vert^{2}$, or small transmit power to user $k$, 
$\left\vert w_{lk} \right\vert^{2}$, at current iteration are given larger weights $\mu_l$ or $\nu_{lk}$ in the next iteration. 
This further decreases $\sum_{k \in \mathcal{K}} \left\vert w_{lk} \right\vert^{2}$ or $\left\vert w_{lk} \right\vert^{2}$ 
in the next iteration, 
and eventually forces BS $l$ toward sleep mode (i.e., $\sum_{k \in \mathcal{K}} \left\vert w_{lk} \right\vert^{2} = 0$) or 
to be removed from user $k$'s serving cluster (i.e., $\left\vert w_{lk} \right\vert^{2} = 0$). 
The weight $\mu_l$ has the effect of putting appropriate BSs to sleep mode, while $\nu_{lk}$ has the effect of 
determining the BS cluster size for user $k$, which in turn affects the backhaul capacity consumption of user $k$.

The resulting optimization problem after the $\ell_1$-norm approximation is formulated as follows:  
%\begin{subequations} 
\begin{align} \label{prob:data_sharing_approx}
\displaystyle \mini_{\left\{w_{lk} \right\}} & \quad \sum_{l\in\mathcal{L}} \sum_{k\in \mathcal{K}} 
\alpha_{lk} \left\vert w_{lk} \right\vert^{2} \\
 \sbto  & \quad \eqref{sinr_const}, ~~ \eqref{Per_BS_Power} \nonumber
\end{align}
%\end{subequations}
where $\alpha_{lk} = \eta_l + \mu_l P_{l, \Delta} + \rho_l \nu_{lk} r_k$. 
Problem \eqref{prob:data_sharing_approx} is a weighted sum transmit power minimization problem, which can be solved 
efficiently through the uplink-downlink duality approach \cite{dahrouj10} or by transforming it into an SOCP problem \cite{wiesel2006}. 
We now summarize the proposed algorithm to solve the total power minimization problem \eqref{prob:data_sharing} for the 
data-sharing strategy in Algorithm~\ref{alg:data_sharing}. 

\begin{algorithm}[t]
{\bf Initialization}: Set the initial values for $\left\{\mu_{l}, \nu_{lk} \right\}$ according to \eqref{wgt_update_data} 
with the $\left\{ w_{lk} \right\}$ chosen as a feasible point of problem \eqref{prob:data_sharing}; \\
{\bf Repeat}:
\begin{enumerate}
\item Fix $\left\{\mu_{l}, \nu_{lk}\right\}$, find the optimal 
$\left\{ w_{lk} \right\}$ by solving problem \eqref{prob:data_sharing_approx} 
using the uplink-downlink duality approach \cite{dahrouj10} or by transforming it into an SOCP problem \cite{wiesel2006};  
\item Update $\left\{\mu_{l}, \nu_{lk} \right\}$ 
according to \eqref{wgt_update_data}. 
\end{enumerate}
{\bf Until} convergence
\caption{Total Power Minimization for Data-Sharing Strategy}
\label{alg:data_sharing}
\end{algorithm}

Note that a similar problem as to \eqref{prob:data_sharing} is considered in our previous work \cite{binbin13}, 
where we formulate the problem as a tradeoff between the BS transmit power and the backhaul capacity. 
This paper considers a more realistic BS power consumption model with sleep mode capability, and also accounts for backhaul power consumption. 
The considered problem \eqref{prob:data_sharing} in this paper 
can also be thought of as providing a tradeoff between the per-BS power consumption and the per-BS 
backhaul capacity consumption, where the tradeoff constant $\rho_l$ is specifically chosen according to the backhaul power consumption model 
\eqref{eq:bkhaul_power}.

%%%% new subsection %%%%%%%%%%%

%{\color{blue}
\subsection{Convergence Analysis}\label{sec:Data_Converge}

Algorithm~\ref{alg:data_sharing} relies on the reweighting heuristic \eqref{wgt_update_data} to deactivate BSs and 
reduce the BS cluster size for energy saving purpose. 
To establish the convergence proof for Algorithm~\ref{alg:data_sharing} under arbitrary reweighting function is challenging, however, 
we show in the following that if the reweighting function is chosen as 
\begin{equation}\label{eq:reweight}
f\left(x, \tau\right) = \frac{1}{\left(x + \tau\right) \ln \left( 1 + \tau^{-1} \right) } ~, 
\end{equation}
i.e. the constants in \eqref{wgt_update_data} are chosen as 
$c_1 = \frac{1}{\ln \left( 1 + \tau_1^{-1} \right)}, c_2 = \frac{1}{\ln \left( 1 + \tau_2^{-1} \right)}$, 
Algorithm~\ref{alg:data_sharing} can be seen as a special case of the MM algorithms \cite{Bharath11} and is guaranteed to converge. 
\begin{theorem}\label{thm:1}
Starting with any initial point, the sequence $\left\{ w_{lk}^{(n)} \right\}_{n=1}^{\infty}$ generated by Algorithm~\ref{alg:data_sharing} with the reweighting function chosen as \eqref{eq:reweight} 
is guaranteed to converge. 
\end{theorem}
\begin{IEEEproof}
See Appendix~\ref{apdx:a}.
\end{IEEEproof}

Finally, we point out that the choice of the reweighting function \eqref{eq:reweight} is not unique. 
There exist other reweighting functions that may work well in different problem setups \cite{Candes08}. 
Recently, \cite{ZhouTaoChen} has experimented with other approximation functions to the $\ell_0$-norm, 
e.g. exponential function and arc-tangent function, in addition to the logarithmic 
function \eqref{lim_approx} used in this paper, and observed similar effectiveness of these functions in inducing sparsity. 

%}

\subsection{Complexity Analysis}

Algorithm~\ref{alg:data_sharing} is an iterative procedure between updating the weights $\left\{\mu_{l}, \nu_{lk}\right\}$ and 
solving the weighted transmit power minimization problem \eqref{prob:data_sharing_approx}. 
The problem \eqref{prob:data_sharing_approx} can be formulated as an SOCP and solved using the 
interior-point method, e.g. using the convex optimization solver \cite{CVX}. 
The total number of variables in problem \eqref{prob:data_sharing_approx} is $LK$ and the total number of SOC 
constraints is $\left(L + K \right)$. The complexity order for solving such a problem through
interior-point method is given as $O\left( \left(L + K \right) \left(LK\right)^3 \right)$ \cite{boyd}. 
Assuming that Algorithm~\ref{alg:data_sharing} requires a total number of $T_1$ weight updates, 
the overall complexity order for Algorithm~\ref{alg:data_sharing} is then $O\left( T_1\left(L + K \right) \left(LK\right)^3\right)$. 

Note that in the above complexity order, $K$ is the number of scheduled users, which is comparable to the number of active BSs in
the network. In addition, instead of considering all the $L$ BSs in the entire network, we can set the nearest $L_c < L$ BSs around each scheduled user as its candidate serving BS cluster. 
This further reduces the computational complexity for Algorithm~\ref{alg:data_sharing} with negligible performance loss.

\subsection{Generalization to the Multi-Antenna System}

Algorithm~\ref{alg:data_sharing} can be readily generalized to the case with multiple transmit antennas at each BS. 
In such case, one only needs to replace the beamforming coefficient $w_{lk}$ with the 
beamforming vector $\mathbf{w}_{lk} \in \mathbb{C}^{N_l \times 1}$ from BS $l$ to user $k$, where $N_l$ is the number of 
antennas at BS $l$. 
The rest of the optimization parameters are straightforward extensions based on $\mathbf{w}_{lk}$ \cite{binbin13}. 

Algorithm~\ref{alg:data_sharing} can also be applied to the case with multiple receive antennas at each user but with fixed 
receive beamformer. In this case, the only change is to replace the channel gain vector $\mathbf{h}_{k}$ 
with the effective channel gain $\tilde{\mathbf{h}}_{k} = \mathbf{H}_{k} \mathbf{u}_{k}$, where 
$\mathbf{H}_{k} \in \mathbb{C}^{L \times M_k}$ and $\mathbf{u}_{k} \in \mathbb{C}^{M_k \times 1}$ are the channel matrix 
seen by user $k$ and the receive beamformer at user $k$, $M_k$ is the number receive antennas at user $k$. 
However, the joint design of transmit beamformer and receive beamformer for the multiple receive antennas case is more complicated.
One possible way is to iteratively design the transmit beamformer assuming fixed receive beamformer and update the receiver as the 
optimal minimum mean square error (MMSE) beamformer.

\begin{figure}[!t]
  \centering
	\psfrag{p}[tc][Bc][1]{Central Processor}
	\psfrag{q}[tc][Bc][0.9]{$\hat{x}_l = w_{l1}s_1 + w_{l2}s_2$}
	\psfrag{l}[tc][Bc][0.9]{$l = 1, 2, 3$}
	\psfrag{a}[tc][tc][0.9]{$x_1$}
	\psfrag{b}[tc][tc][0.9]{$x_2$}
	\psfrag{c}[tc][tc][0.9]{$x_3$}
	\psfrag{i}[tl][tc][0.8]{$R_1^{BH}$}
	\psfrag{j}[tl][tc][0.8]{$R_2^{BH}$}
	\psfrag{k}[tl][tc][0.8]{$R_3^{BH}$}
	\psfrag{x}[tc][cl][0.8]{$x_1 = \hat{x}_1 + e_1$}
	\psfrag{y}[tc][cl][0.8]{$x_2 = \hat{x}_2 + e_2$}
	\psfrag{z}[tc][cl][0.8]{$x_3 = \hat{x}_3 + e_3$}
	\psfrag{m}[tc][tc][0.8]{BS $1$}
	\psfrag{e}[tc][tc][0.8]{BS $2$}
	\psfrag{g}[tc][tc][0.8]{BS $3$}
	\psfrag{n}[tc][tc][0.8]{User $1$}
	\psfrag{f}[tc][tc][0.8]{User $2$}
  \includegraphics[width= 0.45\textwidth]{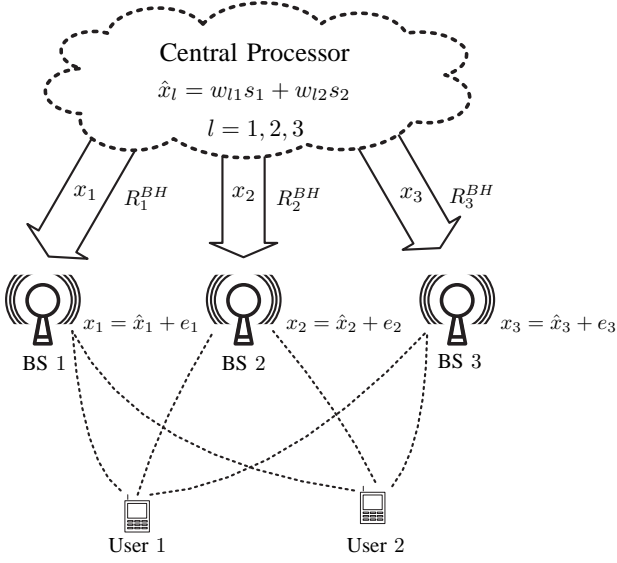}
\caption{Downlink C-RAN with compression transmission strategy. In this illustrative example, the CP centrally precodes 
both users' messages $s_1$ and $s_2$ to $\hat{x}_l, l=1,2,3,$ and forwards the compressed precoded signals 
$x_l,l=1,2,3,$ to each of the BSs. Each BS transmits the compressed beamformed signal received from the CP to both users.}
\label{fig:CompressionModel}
\end{figure}

\section{Compression Strategy}\label{sec:compression}

In this section, we aim to minimize the total power consumption for downlink C-RAN under the 
compression strategy. 
% \eqref{eq:total_power}

\subsection{Problem Formulation}

Consider the compression transmission strategy for downlink C-RAN as illustrated in Fig.~\ref{fig:CompressionModel}.
Let $\hat{x}_l = \sum_{k \in \mathcal{K}} w_{lk} s_k$ denote the beamformed signal formed in the CP for BS $l$. 
The CP compresses $\hat{x}_l$ into $x_l$ and sends $x_l$ to BS $l$. 
In this paper, we assume that each $\hat{x}_l$ is compressed independently\footnote{Correlated compression is also possible and has been considered in \cite{Park13}.} and model the compression procedure as the following forward test channel: 
\begin{equation}\label{eq:compressX}
x_l = \hat{x}_l + e_l, \quad l \in \mathcal{L}, 
\end{equation}
where $e_l \in \mathbb{C}$ is the quantization noise independent of $\hat{x}_l$ 
and is assumed to be Gaussian distributed with zero mean and variance $q_l^{2}$. 
Substituting \eqref{eq:compressX} to \eqref{eq:TxPower}, the transmit power at BS $l$ under the compression strategy can be written as 
\begin{equation}\label{eq:Tx_power_compression}
P_{l, tx} = \sum_{k \in \mathcal{K}} \left\vert w_{lk} \right\vert^{2} + q_l^{2}, \quad l \in \mathcal{L}. 
\end{equation}
Comparing \eqref{eq:Tx_power_compression} with \eqref{eq:Tx_power_data_sharing}, we can see that different from the data-sharing 
strategy, the BS transmit power in the compression strategy involves a quantization noise power in addition to the beamforming power.

Substituting \eqref{eq:compressX} into \eqref{eq:yk_general}, the received signal $y_k$ at user $k$ 
under the compression strategy can be written as 
\begin{equation}\label{eq:y_k_compression}
y_k = \mathbf{h}_k^{H} \mathbf{w}_k s_k + \sum_{j \neq k}\mathbf{h}_k^{H} \mathbf{w}_j s_j +  \mathbf{h}_k^{H} \mathbf{e} + n_k, \quad k \in \mathcal{K}, 
\end{equation}
where $\mathbf{e} = \left[e_1, e_2, \cdots, e_L \right]^{T}$ is the quantization noise vector transmitted from all the $L$ BSs. 
As we can see, besides the inter-user interference and background noise, each user now also receives an additional quantization noise term 
$\mathbf{h}_k^{H} \mathbf{e}$ from the BSs. The user received SINR is expressed as 
\begin{equation}\label{eq:sinr_compression}
		\text{SINR}_k =
\frac{\left\vert\mathbf{h}_k^{H}\mathbf{w}_k \right\vert^2}{\sum_{j \neq
k}\left\vert\mathbf{h}_k^{H}\mathbf{w}_j\right\vert^2 + 
\sum_{l \in \mathcal{L}} \left\vert h_{lk} q_l \right\vert^2 + \sigma^2}, \quad k \in \mathcal{K}.
\end{equation}

The backhaul capacity consumption for the compression strategy is related to the level of quantization noise $q_l^2$: 
lower quantization noise requires higher backhaul rate. 
Under the forward test channel \eqref{eq:compressX}, the achievable compression rate is the mutual information between 
$x$ and $\hat{x}$, according to rate-distortion theory \cite{EIT}, given as 
$\log_2 \left( 1 + \frac{\sum_{k \in \mathcal{K}} \left\vert w_{lk} \right\vert^{2}}{q_l^2}  \right)$,  
where $\sum_{k \in \mathcal{K}} \left\vert w_{lk} \right\vert^{2}$ is the power of the signal to be compressed, i.e. $\hat{x}_l$. 
However, practical quantizer may be far from the theoretically ideal quantizer. 
Similar to \cite{PratikEUSIPCO}, we introduce a notion of gap to rate-distortion limit, denote as $\Gamma_q > 1$, 
to account for the loss due to practical quantizer and formulate the backhaul capacity consumption for BS $l$ as 
\begin{equation}\label{eq:bkhaul_compression}
R_{l}^{BH} = \log_2 \left( 1 + \frac{\Gamma_q \sum_{k \in \mathcal{K}} \left\vert w_{lk} \right\vert^{2}}{q_l^2}  \right), \quad l \in \mathcal{L}. 
\end{equation}

Substituting \eqref{eq:Tx_power_compression} and \eqref{eq:bkhaul_compression} into \eqref{eq:total_power}, the total power 
minimization problem for the compression strategy is formulated as follows
\begin{subequations} \label{prob:compression}
\begin{eqnarray} 
& \hspace{-5mm} \displaystyle \min_{\left\{w_{lk}, q_l\right\}} & \hspace{-1mm} \sum_{l\in\mathcal{L}} 
\Bigg( \eta_l \left( \sum_{k \in \mathcal{K}} \left\vert w_{lk} \right\vert^{2} + q_l^2 \right)+ 
\mathbbm{1}_{\left\{ \sum_{k \in \mathcal{K}} \left\vert w_{lk} \right\vert^{2} + q_l^2 \right\}} P_{l, \Delta} \nonumber \\
 & \quad &\hspace{1.2cm} + \rho_l \log_2 \left( 1 + \frac{\Gamma_q \sum_{k \in \mathcal{K}} \left\vert w_{lk} \right\vert^{2}}{q_l^2}  \right) \Bigg) 
\label{obj_compression} \\
& \hspace{-5mm} \st &  \hspace{-1mm} \text{SINR}_k \geq \gamma_k, \quad k \in \mathcal{K} \label{sinr_const_compression} \\
&  \quad & \hspace{-1mm} \sum_{k \in \mathcal{K}} \left\vert w_{lk} \right\vert^{2} + q_l^2 \leq P_l, \quad l \in \mathcal{L} \label{compression_power_const}
\end{eqnarray}
\end{subequations} 
where the SINR in \eqref{sinr_const_compression} is defined in \eqref{eq:sinr_compression}. 
Due to the indicator function as well as the backhaul rate expression in \eqref{obj_compression}, 
the optimization problem \eqref{prob:compression} is nonconvex. 
In the following, we describe the techniques to approximate \eqref{obj_compression} in a convex form.

\subsection{Proposed Algorithm}

The difficulties in solving problem \eqref{prob:compression} lie in both the indicator function and the nonconvex backhaul rate 
expression in the objective function \eqref{obj_compression}. 
For the indicator function, we can utilize the similar technique used in the previous section to approximate 
it using reweighted $\ell_1$-norm: 
\begin{align} \label{eq:rewgt_compression}
\mathbbm{1}_{\left\{ \sum_{k \in \mathcal{K}} \left\vert w_{lk} \right\vert^{2} + q_l^2 \right\}}
 & = \left\Vert \sum_{k \in \mathcal{K}} \left\vert w_{lk} \right\vert^{2} + q_l^2   \right\Vert_0  \nonumber \\ 
 &\approx \beta_l \left( \sum_{k \in \mathcal{K}} \left\vert w_{lk} \right\vert^{2} + q_l^2  \right) 
\end{align}
where $\beta_l$ is iteratively updated according to the following reweighting function
\begin{equation}\label{wgt_mu}
\beta_l = f\left(\sum_{k \in \mathcal{K}} \left\vert w_{lk} \right\vert^{2} + q_l^2 , \tau_3\right) 
= \frac{c_3}{\sum_{k \in \mathcal{K}} \left\vert w_{lk} \right\vert^{2} + q_l^2 + \tau_3}
\end{equation}
where $\left\{w_{lk}, q_l \right\}$ come from the previous iteration, $\tau_3 > 0$ is some constant regularization factor, 
and $c_3$ is a constant. 
%\begin{equation}\label{wgt_mu}
%\beta_l = \frac{1}{\sum_{k \in \mathcal{K}} \left\vert w_{lk} \right\vert^{2} + q_l^2 + \tau_3}
%\end{equation}

For the backhaul rate \eqref{eq:bkhaul_compression}, we can express it as a difference of two logarithmic functions: 
$\log_2 \left( q_l^2 + \Gamma_q \sum_{k \in \mathcal{K}} \left\vert w_{lk} \right\vert^{2}  \right) - 2 \rho_l \log_2 q_l$. 
Although the second term $- 2 \rho_l \log_2 q_l$ is convex in $q_l$, the first term is still nonconvex. 
To deal with the nonconvexity of the backhaul rate, we propose to 
successively approximate the first logarithmic function using the following inequality
\begin{align} \label{ineq:compression}
 &\log_2 \left( q_l^2 + \Gamma_q \sum_{k \in \mathcal{K}} \left\vert w_{lk} \right\vert^{2}  \right) \nonumber \\
 & \leq  ~ \log_2 \lambda_l  +  \frac{ q_l^2 + \Gamma_q \sum_{k \in \mathcal{K}} \left\vert w_{lk} \right\vert^{2}}{\lambda_l \ln 2} - 
\frac{1}{\ln 2} 
\end{align}
due to the concavity of $\log_2(x)$. The above inequality achieves equality if and only if 
\begin{equation}\label{lambda_update}
\lambda_l = q_l^2 + \Gamma_q \sum_{k \in \mathcal{K}} \left\vert w_{lk} \right\vert^{2}. 
\end{equation} 
The right-hand side of \eqref{ineq:compression} is a convex quadratic function in $\left\{ w_{lk}, q_l \right\}$ for fixed 
$\lambda_l$. 
This fact motivates us to successively solve the 
problem \eqref{prob:compression} with $\log_2 \left( q_l^2 + \Gamma_q \sum_{k \in \mathcal{K}} \left\vert w_{lk} \right\vert^{2}  \right)$ replaced by the right-hand side of \eqref{ineq:compression} for fixed $\lambda_l$, then to 
iteratively update $\lambda_l$ according to \eqref{lambda_update}. 

Combining the above described $\ell_1$-norm reweighting and successive convex approximation techniques, 
we get the resulting optimization problem under fixed $\beta_l$ and $\lambda_l$ as 
%\begin{subequations} 
\begin{align} \label{prob:approx_compression1}
\displaystyle \mini_{\left\{w_{lk}, q_l\right\}} & \quad \sum_{l\in\mathcal{L}} \sum_{k\in\mathcal{K}} \phi_{l} \left\vert w_{lk} \right\vert^{2} + \sum_{l\in\mathcal{L}} \left( \psi_{l} q_l^2 - 2 \rho_l \log_2 q_l\right) \nonumber \\ 
& \hspace{1.5cm} + \underbrace{\sum_{l\in\mathcal{L}} \rho_l \left( \log_2 \lambda_l -  \frac{1}{\ln 2} \right)}_{\text{constant}} \\
 \sbto  & \quad \eqref{sinr_const_compression}, ~~ \eqref{compression_power_const} \nonumber
\end{align}
%\end{subequations}
where $\phi_{l} = \eta_l + \beta_l P_{l, \Delta} + \frac{\rho_l \Gamma_q}{\lambda_l \ln 2}$ and 
$\psi_{l} = \eta_l + \beta_l P_{l, \Delta} + \frac{\rho_l}{\lambda_l \ln 2}$. 
Similar to the SINR constraint in \eqref{sinr_const}, the constraint \eqref{sinr_const_compression} can also be 
equivalently reformulated as an SOC constraint. 
Thus, problem \eqref{prob:approx_compression1} is a convex optimization problem and can be solved 
efficiently using standard convex optimization solver, e.g. \cite{CVX}, with polynomial complexity.

\begin{algorithm}[t]
{\bf Initialization}: Set the initial values for $\left\{ \beta_l \right\}$ and $\left\{ \lambda_l \right\}$
according to \eqref{wgt_mu} and \eqref{lambda_update} respectively with $\left\{ w_{lk}, q_l \right\}$ chosen as a feasible 
point of problem \eqref{prob:compression}; \\
{\bf Repeat}:
\begin{enumerate}
\item Fix $\left\{ \beta_l, \lambda_l \right\}$, find the optimal 
$\left\{ w_{lk}, q_l \right\}$ by solving the convex optimization problem \eqref{prob:approx_compression1}; 
\item Update $\left\{ \beta_l \right\}$ and $\left\{ \lambda_l \right\}$
according to \eqref{wgt_mu} and \eqref{lambda_update} respectively. 
\end{enumerate}
{\bf Until} convergence
\caption{Total Power Minimization for Compression Strategy}
\label{alg:compression}
\end{algorithm}

%{\color{blue} 
We summarize the proposed algorithm for solving problem \eqref{prob:compression} in 
Algorithm~\ref{alg:compression}, which admits guaranteed convergence property as stated in the following theorem. 
\begin{theorem}\label{thm:2}
Starting with any initial point, the sequence $\left\{ w_{lk}^{(n)}, q_l^{(n)} \right\}_{n=1}^{\infty}$ generated by Algorithm~\ref{alg:compression} with the reweighting function in \eqref{wgt_mu} chosen as 
\eqref{eq:reweight} is guaranteed to converge. 
\end{theorem}
\begin{IEEEproof}
See Appdendix~\ref{apdx:b}
\end{IEEEproof}
%}

Algorithm~\ref{alg:compression} shows a similar computational complexity as Algorithm~\ref{alg:data_sharing} but with 
additional $L$ quantization noise variables to be optimized in each iteration.  
Assuming that Algorithm~\ref{alg:compression} converges in $T_2$ iterations, its complexity order 
is then given as $O\left( T_2\left(L + K \right) \left(LK + L\right)^3 \right)$. 

%Due to the two-loop nature of Algorithm~\ref{alg:compression}, a coarse solution to problem \eqref{prob:approx_compression} 
%through Algorithm~\ref{alg:SCA} in the inner loop is typically 
%already sufficient for the updates of $\beta_l$ in the outer loop. 
%In practice, we update $\beta_l$ 
%whenever the objective value of problem \eqref{prob:approx_compression} has 
%converged to within some fixed tolerance (e.g., 1\% from one iteration to the next). 
%This reduces the number of updates $T_2$ required for $\lambda_l$. 
%Moreover, although Algorithm~\ref{alg:SCA} guarantees to converge under any random initialization of $\lambda_l$, 
%in simulations we initialize  
%$\lambda_l$ according to \eqref{lambda_update} with the $\left\{ w_{lk}, q_l \right\}$ from previous iteration. 
%This further reduces the total number of updates needed for Algorithm~\ref{alg:compression} to converge. 

\subsection{Generalization to the Multi-Antenna System}

Algorithm~\ref{alg:compression} can be readily applied to the scenario where multiple transmit antennas are available at the BSs assuming 
that the CP performs independent compression for each antenna. 
Joint compression among the antennas may improve the performance but results in a different optimization problem. 
Similar to Algorithm~\ref{alg:data_sharing}, 
generalization of Algorithm~\ref{alg:compression} to multiple receive antennas at the user side is also straightforward if the receive beamformer is assumed to be fixed, however, joint design of transmit and receive beamformer is 
by no means trivial and requires additional efforts.

\section{Numerical Evaluation of Energy Efficiency}\label{sec:simulations}

\begin{table}[t]
\centering
\caption{Simulation Parameters.}
\label{table:system-parameter}
\begin{tabular}{|c|c|}
\hline 
Cellular  & Hexagonal  \\
     Layout        &  $7$-cell wrapped-around \\\hline
Channel bandwidth & $10$ MHz    \\ \hline
Distance between cells  &  $0.8$ km \\ \hline
Number of RRHs$/$cell  &  $4$   \\ \hline
Number of antennas$/$(RRH, user)  &   $(1, 1)$ \\ \hline
Maximum transmit power for RRH $P_l$   &  $20$ Watts \\ \hline
Active mode power for RRH $P_{l, active}$ & $84$ Watts \\ \hline
Sleep mode power for RRH $P_{l, sleep}$ & $56$ Watts \\ \hline
Slope of transmit power $\eta_l$ & $2.8$ \\ \hline
Backhaul link capacity $C_l$ & $100$ Mbps \\ \hline
Maximum backhaul power $P_{l,max}^{BH}$ & $50$ Watts \\ \hline
 Antenna gain & $15$ dBi \\ \hline
 Background noise  & $-169$ dBm/Hz \\ \hline
 Path loss from RRH to user & $128.1+ 37.6 \log_{10}(d)$ \\ \hline
Log-normal shadowing & $8$ dB \\ \hline
Rayleigh small scale fading & $0$ dB  \\ \hline
SNR gap $\Gamma_m$ & $0$ dB \\ \hline
Gap to rate-distortion limit $\Gamma_q$ & $4.3$ dB \\ \hline
Reweighting parameters ($\tau_1, \tau_2, \tau_3$) & $(10^{-5}, 10^{-8}, 10^{-5})$ \\ \hline
\end{tabular}
\end{table}

\begin{figure}[t]
\centering
\psfrag{xlabel}[cc][cc][0.8]{km}
\psfrag{ylabel}[cc][cc][0.8]{km}
\psfrag{0}[cr][cr][0.7]{$0$}
\psfrag{0.5}[cr][cr][0.7]{$0.5$}
\psfrag{1}[cr][cr][0.7]{$1$}
\psfrag{1.5}[cr][cr][0.7]{$1.5$}
\psfrag{-0.5}[cr][cr][0.7]{$-0.5$}
\psfrag{-1}[cr][cr][0.7]{$-1$}
\psfrag{-1.5}[cc][cc][0.7]{$-1.5$}
\psfrag{x4}[cr][cr][0.7]{$0$}
\psfrag{x5}[cr][cr][0.7]{$0.5$}
\psfrag{x6}[cr][cr][0.7]{$1$}
\psfrag{x7}[cr][cr][0.7]{$1.5$}
\psfrag{x3}[cr][cr][0.7]{$-0.5$}
\psfrag{x2}[cr][cr][0.7]{$-1$}
\psfrag{x1}[cc][cc][0.7]{$-1.5$}

  \centering
  \includegraphics[width=0.45\textwidth]{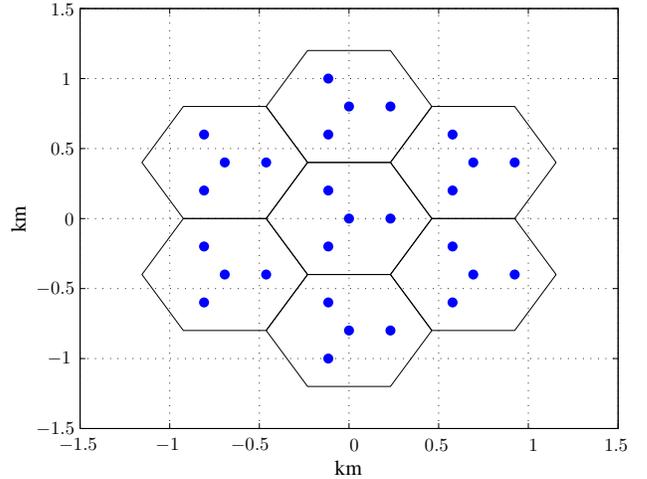}
\caption{A cellular topology with $7$ cells and $4$ RRHs each cell, where each dot represents a RRH.}
\label{fig:topology}
\end{figure}

In this section, we evaluate the energy efficiency of the data-sharing and compression strategies 
for downlink C-RAN using the proposed algorithms for a $7$-cell network with wrapped-around topology.
Each cell here refers to a geographic area with $4$ RRHs as shown in Fig.~\ref{fig:topology}. 
Equivalently, the network consists of $28$ BSs. 
The out-of-cell interference combined with background noise is set as $-150$ dBm$/$Hz. 
The gap to rate-distortion limit is set as $\Gamma_q = 4.3$ dB corresponding to the uncoded fixed rate 
uniform scalar quantizer \cite{Gray98}. 
For simplicity, the SNR gap is set to be $\Gamma_m = 0$ dB. 
The parameters in the BS power consumption model \eqref{eq:bs_power} are taken from 
\cite{Auer11} while the parameters for the backhaul power model \eqref{eq:bkhaul_power} are from \cite{Fehske10}. 
All the parameters related to the simulations are listed in Table~\ref{table:system-parameter}.

\begin{figure}[t]
\centering
\psfrag{xlabel}[tc][Bc][0.8]{Iteration }
\psfrag{ylabel}[Bc][tc][0.8]{Number of Active BSs}
\psfrag{Algorithm 1 w/ 1 user/cell extraaaaaa}[Bl][Bl][0.65]{Data-Sharing (Alg.1) w/ $1$ user$/$cell}
\psfrag{Algorithm 1 w/ 2 users/cell}[Bl][Bl][0.65]{Data-Sharing (Alg.1) w/ $2$ users$/$cell}
\psfrag{Algorithm 1 w/ 3 users/cell}[Bl][Bl][0.65]{Data-Sharing (Alg.1) w/ $3$ users$/$cell}
\psfrag{Algorithm 2 w/ 1 user/cell}[Bl][Bl][0.65]{Compression (Alg.2) w/ $1$ user$/$cell}
\psfrag{Algorithm 2 w/ 2 users/cell}[Bl][Bl][0.65]{Compression (Alg.2) w/ $2$ users$/$cell}
\psfrag{Algorithm 2 w/ 3 users/cell}[Bl][Bl][0.65]{Compression (Alg.2) w/ $3$ users$/$cell}
\psfrag{x1}[cc][cr][0.7]{$10$}
\psfrag{x2}[cc][cc][0.7]{$20$}
\psfrag{x3}[cc][cc][0.7]{$30$}
\psfrag{x4}[cc][cc][0.7]{$40$}
\psfrag{x5}[cc][cc][0.7]{$50$}
\psfrag{x6}[cc][cc][0.7]{$60$}
\psfrag{x7}[cc][cc][0.7]{$70$}
\psfrag{y0}[rc][rc][0.7]{$5$}
\psfrag{y1}[rc][rc][0.7]{$10$}
\psfrag{y2}[rc][rc][0.7]{$15$}
\psfrag{y3}[rc][rc][0.7]{$20$}
\psfrag{y4}[rc][rc][0.7]{$25$}
\psfrag{y5}[rc][rc][0.7]{$30$}
\psfrag{y6}[rc][rc][0.7]{$35$}
  \centering
  \includegraphics[width=0.45\textwidth]{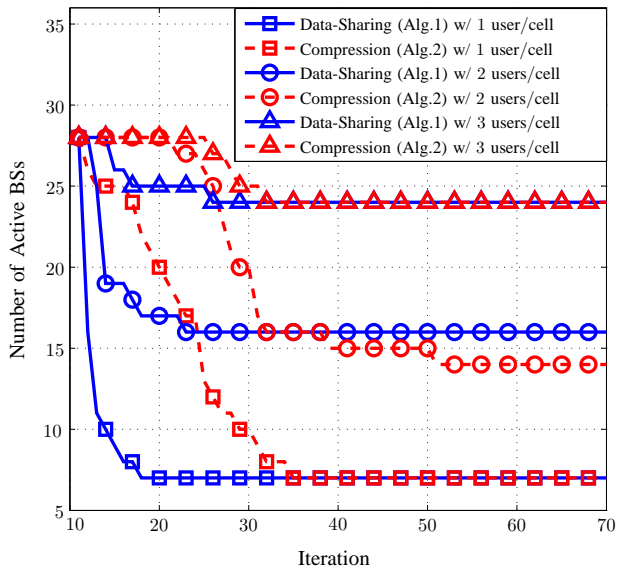}
\caption{Trajectories of number of active BSs under $r_k = 20$ Mbps target rate for each user.}
\label{fig:DataSharingActiveBSs}
\end{figure}

\begin{figure}[t]
\centering
\psfrag{xlabel}[tc][Bc][0.8]{Iteration}
\psfrag{ylabel}[Bc][tc][0.7]{Objective Value of Problem \eqref{prob:data_sharing_approx_lim}$/$\eqref{prob:compression_approx}}
\psfrag{Algorithm 1 w/ 1 user/cell extraaaaaaaa}[Bl][Bl][0.65]{Data-Sharing (Alg.1) w/ $1$ user$/$cell}
\psfrag{Algorithm 1 w/ 2 users/cell}[Bl][Bl][0.65]{Data-Sharing (Alg.1) w/ $2$ users$/$cell}
\psfrag{Algorithm 1 w/ 3 users/cell}[Bl][Bl][0.65]{Data-Sharing (Alg.1) w/ $3$ users$/$cell}
\psfrag{Algorithm 2 w/ 1 user/cell}[Bl][Bl][0.65]{Compression (Alg.2) w/ $1$ user$/$cell}
\psfrag{Algorithm 2 w/ 2 users/cell}[Bl][Bl][0.65]{Compression (Alg.2) w/ $2$ users$/$cell}
\psfrag{Algorithm 2 w/ 3 users/cell}[Bl][Bl][0.65]{Compression (Alg.2) w/ $3$ users$/$cell}
\psfrag{x1}[cc][cr][0.7]{$10$}
\psfrag{x2}[cc][cc][0.7]{$20$}
\psfrag{x3}[cc][cc][0.7]{$30$}
\psfrag{x4}[cc][cc][0.7]{$40$}
\psfrag{x5}[cc][cc][0.7]{$50$}
\psfrag{x6}[cc][cc][0.7]{$60$}
\psfrag{x7}[cc][cc][0.7]{$70$}
\psfrag{x8}[cc][cc][0.7]{$80$}
\psfrag{0}[rc][rc][0.7]{$0$}
\psfrag{400}[rc][rc][0.7]{$400$}
\psfrag{800}[rc][rc][0.7]{$800$}
\psfrag{1200}[rc][rc][0.7]{$1200$}
\psfrag{1600}[rc][rc][0.7]{$1600$}
\psfrag{2000}[rc][rc][0.7]{$2000$}
  \includegraphics[width=0.45\textwidth]{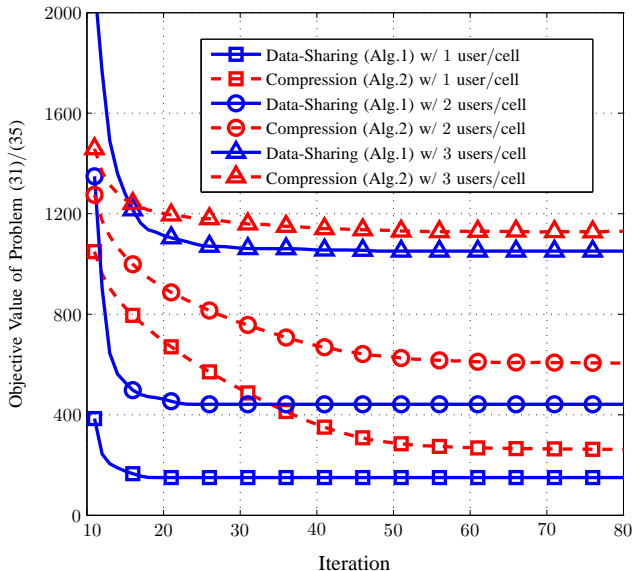}
\caption{Convergence behavior of proposed algorithms under $r_k = 20$ Mbps target rate for each user.}
\label{fig:compressionObj}
\end{figure}

We first evaluate the effectiveness of the proposed $\ell_1$-norm reweighting technique in turning off BSs.
We plot the number of active BSs remained in each iteration for both data-sharing (Algorithm~\ref{alg:data_sharing}) 
and compression (Algorithm~\ref{alg:compression}) strategies in Fig.~\ref{fig:DataSharingActiveBSs}. 
The user target rate is set to be $20$ Mbps for every user and different number of scheduled users are simulated. %{\color{blue}
Instead of considering all the $28$ BSs in the entire network as potential serving BSs for each user, we set the 
initial BS cluster for each user as the strongest $L_c = 14$ BSs to 
reduce the amount of CSI acquisition for the CP. %}
From Fig.~\ref{fig:DataSharingActiveBSs} we can see that all the BSs are active at the first iteration, however,  
the number of active BSs decreases as the iteration goes on.  
Intuitively, more users are served, more BSs need to remain active. 
This can be verified from Fig.~\ref{fig:DataSharingActiveBSs}. 
We also observe from Fig.~\ref{fig:DataSharingActiveBSs} that Algorithm~\ref{alg:data_sharing} for data-sharing 
exhibits faster convergence speed than Algorithm~\ref{alg:compression} for compression, 
where the former converges within $30$ iterations while the later requires $50$ iterations to converge 
in the worst case.

%{\color{blue}
We then evaluate the convergence behavior of the proposed algorithms in Fig.~\ref{fig:compressionObj}.  
As in Fig.~\ref{fig:DataSharingActiveBSs}, different number of scheduled users are tested and each user's 
target rate is set to be $20$ Mbps. 
As we can see, the objective values monotonically decrease and converge for both the data-sharing (Algorithm~\ref{alg:data_sharing}) 
and the compression (Algorithm~\ref{alg:compression}) strategies. 
Similar to Fig.~\ref{fig:DataSharingActiveBSs}, we also observe faster convergence speed for data-sharing than 
for compression in Fig.~\ref{fig:compressionObj}. 
This is possibly due to the fact that compression strategy involves more variables to be optimized. 
%}

\begin{figure*}[t]
\centering
\psfrag{xlabel}[tc][cc][0.7]{User Target Rate (Mbps)}
\psfrag{ylabel}[tc][cc][0.7]{Total Power Consumption (kWatts)}
\psfrag{Compression}[Bl][Bl][0.65]{Compression}
\psfrag{Data Sharing}[Bl][Bl][0.65]{Data-Sharing}
\psfrag{Per Cell CoMP}[Bl][Bl][0.65]{Per-Cell CoMP}
\psfrag{Single BS Association}[Bl][Bl][0.65]{Single BS Association}
\psfrag{x1}[cc][cc][0.7]{\hspace{2mm}$10$}
\psfrag{x2}[cc][cc][0.7]{$20$}
\psfrag{x3}[cc][cc][0.7]{$30$}
\psfrag{x4}[cc][cc][0.7]{$40$}
\psfrag{x5}[cc][cc][0.7]{$50$}
\psfrag{x6}[cc][cc][0.7]{$60$}
\psfrag{x7}[cc][cc][0.7]{$70$}
\psfrag{1500}[cc][cc][0.7]{\hspace{2mm}$1.5$}
\psfrag{2000}[cc][cc][0.7]{\hspace{2mm}$2.0$}
\psfrag{2500}[cc][cc][0.7]{\hspace{2mm}$2.5$}
\psfrag{3000}[cc][cc][0.7]{\hspace{2mm}$3.0$}
\psfrag{3500}[cc][cc][0.7]{\hspace{2mm}$3.5$}
\psfrag{4000}[cc][cc][0.7]{\hspace{2mm}$4.0$}
\psfrag{4500}[cc][cc][0.7]{\hspace{2mm}$4.5$}
\psfrag{5000}[cc][cc][0.7]{\hspace{2mm}$5.0$}
\subfloat[1 user per cell]{\includegraphics*[width=0.33\textwidth]{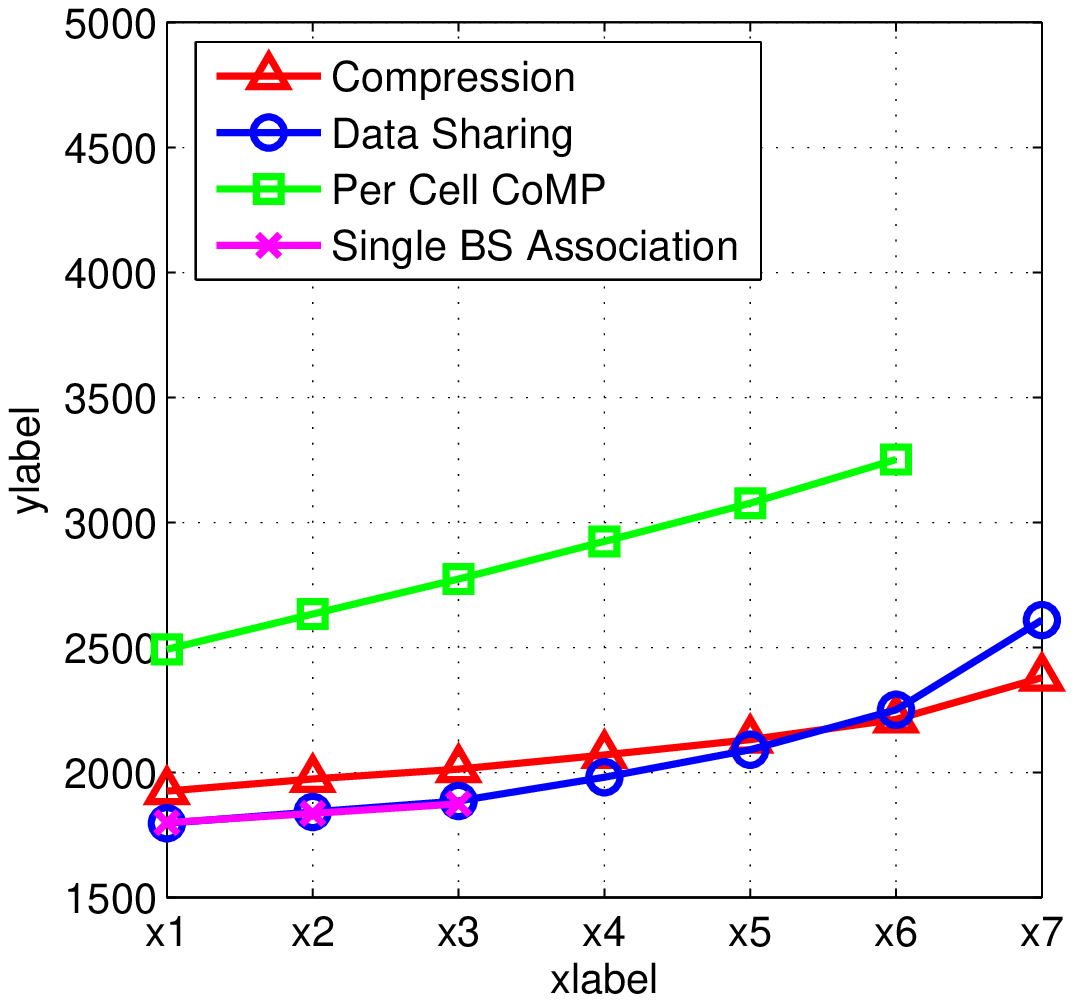}%
\label{sim:power_u1}}
\hfil
\subfloat[2 users per cell]{\includegraphics*[width=0.33\textwidth]{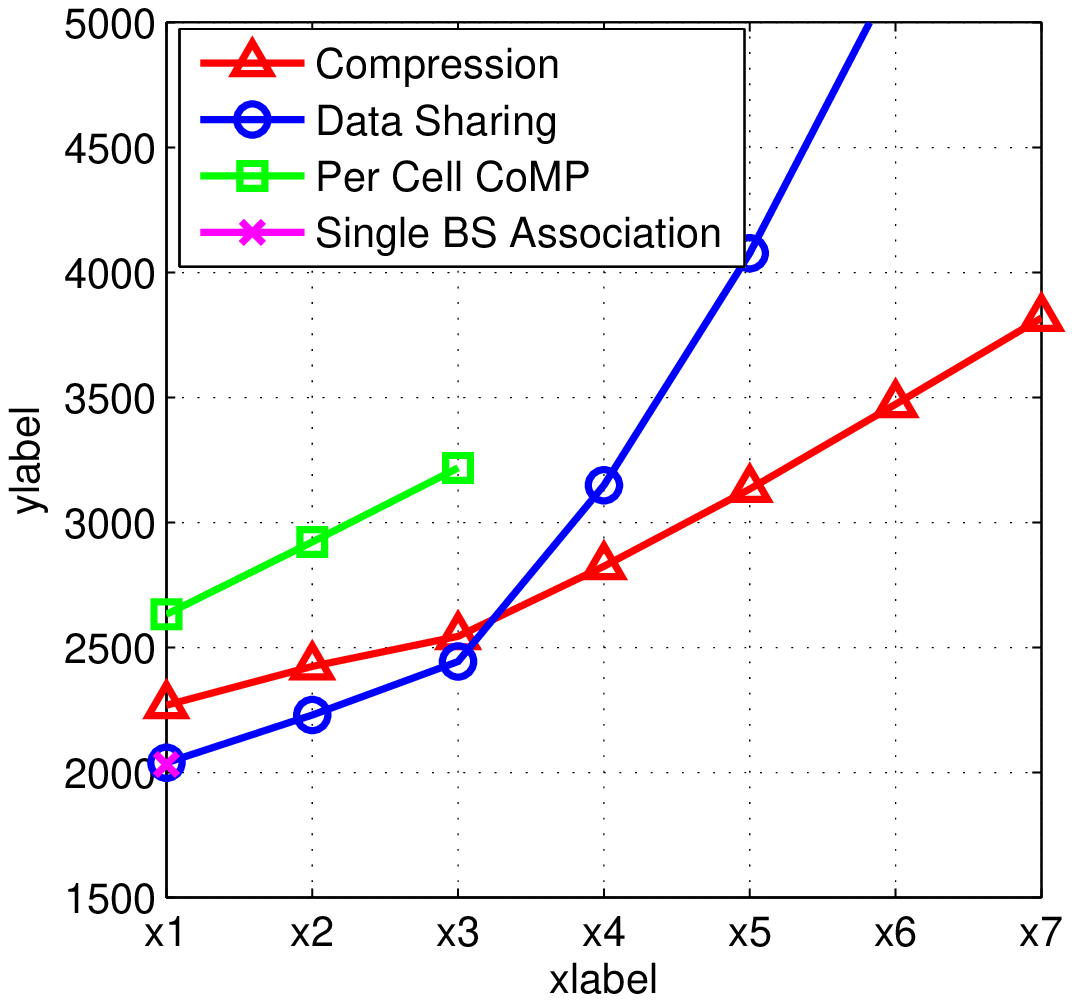}%
\label{sim:power_u2}}
\hfil
\subfloat[3 users per cell]{\includegraphics*[width=0.33\textwidth]{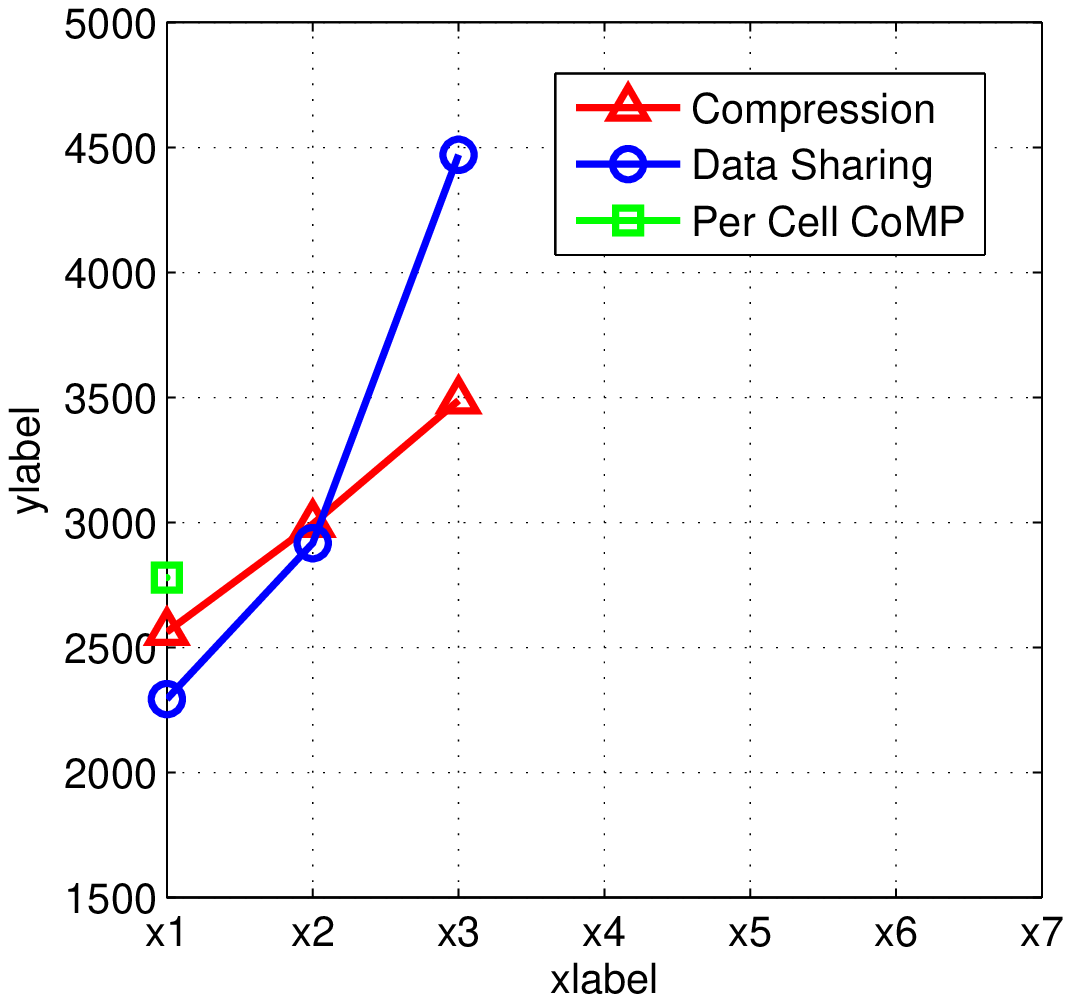}%
\label{sim:power_u3}}
\caption{Total power consumption comparison between different schemes. 
Missing points indicate that the corresponding scheme is infeasible. 
For instance, ``Single BS Association'' is feasible only for the first 
three points in (a), and is only feasible for the very first point in (b), and is infeasible in (c). 
For ``Per-Cell CoMP'' scheme in (c), it is only feasible at $10$ Mbps target rate.}
\label{sim:total_power}
\end{figure*}

We now compare the performance of the data-sharing strategy and the compression strategy in terms of power 
saving in Fig.~\ref{sim:total_power}. %{\color{blue}
In addition, we consider two reference schemes. 
In the first scheme, each user is only served by its strongest BS that is not already associated with another user.
This scheme is termed as ``Single BS Association'', for which the transmit power for each user can be minimized using  
the strategy in \cite{yates1995}. 
In the second scheme, each user's message is shared among the $4$ RRHs in its own cell and 
is cooperatively served by the $4$ RRHs using the 
coordinated beamforming strategy of \cite{dahrouj10}. Such scheme is termed as ``Per-Cell CoMP''.  
The ``Single BS Association'' and ``Per-Cell CoMP'' are two extreme cases in terms of number of active BSs: 
the former only has $K$ (number of scheduled users) active BSs while in the latter all the $28$ BSs in the entire network remain active.%} 
Each point in Fig.~\ref{sim:total_power} is averaged over $100$ channel realizations.

As we can see from Fig.~\ref{sim:total_power}, ``Per-Cell CoMP'' consumes the most power since all the BSs are active in this scheme. 
``Single BS Association'' consumes the least power, similar to the data-sharing strategy, but only at low user rate regime 
because the minimum number of BSs are selected to serve the users in this scheme. 
However, as the user rate or the number of scheduled users increases, ``Single BS Association'' becomes infeasible very quickly. 
For instance, in Fig.~\ref{sim:total_power}(b) where there are $2$ users per cell, ``Single BS Association'' can only support each user 
with $10$ Mbps service rate, while in Fig.~\ref{sim:total_power}(c) the case of 3 users per cell, ``Single BS Association'' 
is not feasible even at $10$ Mbps per user.

It is worth noting that ``Single BS Association'' consumes the same amount of power as data-sharing 
in the low user target rate regime in Fig.~\ref{sim:total_power}, as the latter essentially reduces to single BS association at low user rates. However, there is still significant advantage in migrating signal processing to the cloud in a C-RAN as compared to the conventional cellular architecture in term of computation power saving, which is not included in the model in this paper. It is also worth noting that the optimized data-sharing and compression strategies outperform the non-optimized per-Cell CoMP significantly in Fig.~\ref{sim:total_power}, highlighting the importance of optimization approaches proposed in this paper. 
Other optimized CoMP schemes with larger cooperation cluster may consume less BS transmission power, 
however, the overall power consumptions can be still very high if all the BSs remain active.

We also observe from Fig.~\ref{sim:total_power} that neither the data-sharing nor the compression strategy 
dominates the other over the entire user target rate regime. 
For example in Fig.~\ref{sim:total_power}(b), although data-sharing consumes less power 
than compression when the user target rate is below $30$ Mbps, 
its power consumption increases dramatically with the user rate and 
eventually crosses over the total power consumed by compression after $40$ Mbps target rate. 
Similar trend can be observed from Fig.~\ref{sim:total_power}(a) and \ref{sim:total_power}(c). %{\color{blue}
This trend is parallel to the observation made in \cite{PratikEUSIPCO}, in which these two downlink strategies are 
compared from the utility maximization perspective with limited backhaul constraint. 
It is observed in \cite{PratikEUSIPCO} that with low backhaul rate, data-sharing produces higher utility than compression while 
with high backhaul rate, compression outperforms data-sharing. %}

%It is interesting to note from Fig.~\ref{sim:total_power} that the crossing points of data-sharing strategy and 
%compression strategy in Fig.~\ref{sim:total_power}(a) - \ref{sim:total_power}(c) are all at about $60$ Mbps per-cell sum rate. 
%It appears that $60$ Mbps is a per-cell sum rate threshold for preferring data-sharing strategy or compression strategy 
%in terms of power consumption for the simulated network. 

\begin{figure}[t]
  \centering
	\psfrag{xlabel}[tc][Bc][0.7]{User Target Rate (Mbps) }
\psfrag{ylabel}[Bc][tc][0.7]{Power (kWatts)}
\psfrag{Data Sharing BS Power}[Bl][Bl][0.7]{Data-Sharing BS Power}
\psfrag{Data Sharing Backhaul Power}[Bl][Bl][0.7]{Data-Sharing Backhaul Power}
\psfrag{Compression BS Power}[Bl][Bl][0.7]{Compression BS Power}
\psfrag{Compression Backhaul Power}[Bl][Bl][0.7]{Compression Backhaul Power}
\psfrag{x1}[cc][cc][0.85]{$10$}
\psfrag{x2}[cc][cc][0.85]{$20$}
\psfrag{x3}[cc][cc][0.85]{$30$}
\psfrag{x4}[cc][cc][0.85]{$40$}
\psfrag{x5}[cc][cc][0.85]{$50$}
\psfrag{x6}[cc][cc][0.85]{$60$}
\psfrag{x7}[cc][cc][0.85]{$70$}
\psfrag{1}[cc][cc][0.85]{$1$}
\psfrag{2}[cc][cc][0.85]{$2$}
\psfrag{3}[cc][cc][0.85]{$3$}
\psfrag{4}[cc][cc][0.85]{$4$}
\psfrag{5}[cc][cc][0.85]{$5$}
\psfrag{6}[cc][cc][0.85]{$6$}
\psfrag{7}[cc][cc][0.85]{$7$}	
\psfrag{0}[cc][cc][0.85]{$0$}	
  \includegraphics[width= 0.45\textwidth]{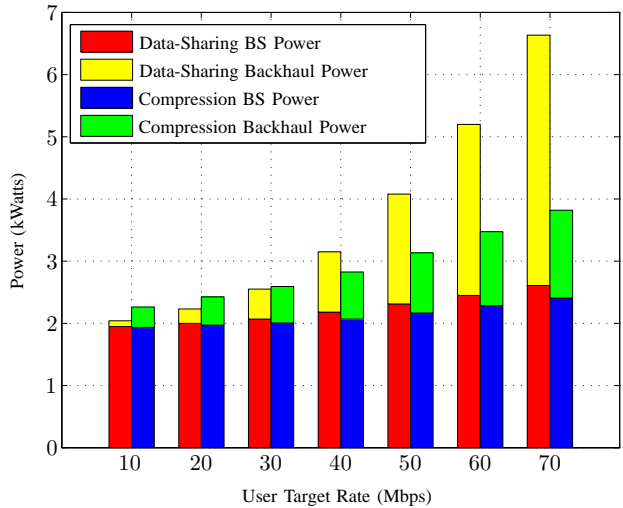}
\caption{Power consumptions of BSs and backhaul links with $2$ users each cell.}
\label{fig:power_decomp}
\end{figure}

To investigate further, 
we plot the individual power consumption of BSs and backhaul links for both the 
data-sharing and the compression strategies in Fig.~\ref{fig:power_decomp} for the case of $2$ users per cell. 
As seen from Fig.~\ref{fig:power_decomp}, although the BS power consumptions for data-sharing and compression 
are similar in each case of the user target rate, the backhaul power consumptions are significantly different 
and are the determining factor in the choice of strategies.  
As the user target rate increases, the backhaul power consumption for data-sharing increases significantly 
and crosses over the compression strategy at around $30$ Mbps. 
This is because in data-sharing, each user's message needs to be delivered to each one of its serving BSs through backhaul links.
So, the backhaul rate directly depends on both the user target rate and the BS cluster size. 
Note that as user target rate increases, the size of serving BS cluster also increases. 
The two factors together contribute to a much higher backhaul rate. 
In contrast, the backhaul rate of compression strategy depends on 
the logarithm of the signal-to-quantization-noise ratio, which only increases gradually as user target rate increases. 
Also, note that in the low user rate regime, 
data-sharing consumes less backhaul power than compression in Fig.~\ref{fig:power_decomp}. 
This is because it is more efficient to share data directly than to compress when only a few BSs are involved.

\begin{figure*}[!t]
\centering
\psfrag{xlabel}[tc][cc][0.6]{User Target Rate (Mbps)}
\psfrag{ylabel}[cc][cc][0.6]{{Fraction of Active BSs}}
\psfrag{Compression}[Bl][Bl][0.65]{Compression}
\psfrag{Data Sharing}[Bl][Bl][0.65]{Data Sharing}
\psfrag{10}[cc][cr][0.7]{$10$}
\psfrag{20}[cc][cc][0.7]{$20$}
\psfrag{30}[cc][cc][0.7]{$30$}
\psfrag{40}[cc][cc][0.7]{$40$}
\psfrag{50}[cc][cc][0.7]{$50$}
\psfrag{60}[cc][cc][0.7]{$60$}
\psfrag{70}[cc][cc][0.7]{$70$}
\psfrag{0.2}[cc][cc][0.7]{$0.2$}
\psfrag{0.3}[cc][cc][0.7]{$0.3$}
\psfrag{0.4}[cc][cc][0.7]{$0.4$}
\psfrag{0.5}[cc][cc][0.7]{$0.5$}
\psfrag{0.6}[cc][cc][0.7]{$0.6$}
\psfrag{0.7}[cc][cc][0.7]{$0.7$}
\psfrag{0.8}[cc][cc][0.7]{$0.8$}
\psfrag{0.9}[cc][cc][0.7]{$0.9$}
\psfrag{0.65}[cc][cc][0.7]{$0.65$}
\psfrag{0.75}[cc][cc][0.7]{$0.75$}
\psfrag{0.85}[cc][cc][0.7]{$0.85$}
\psfrag{0.95}[cc][cc][0.7]{$0.95$}
\psfrag{1}[cc][cl][0.8]{\hspace{-2mm}$1.0$}
\subfloat[1 user per cell]{\includegraphics[width=0.33\textwidth]{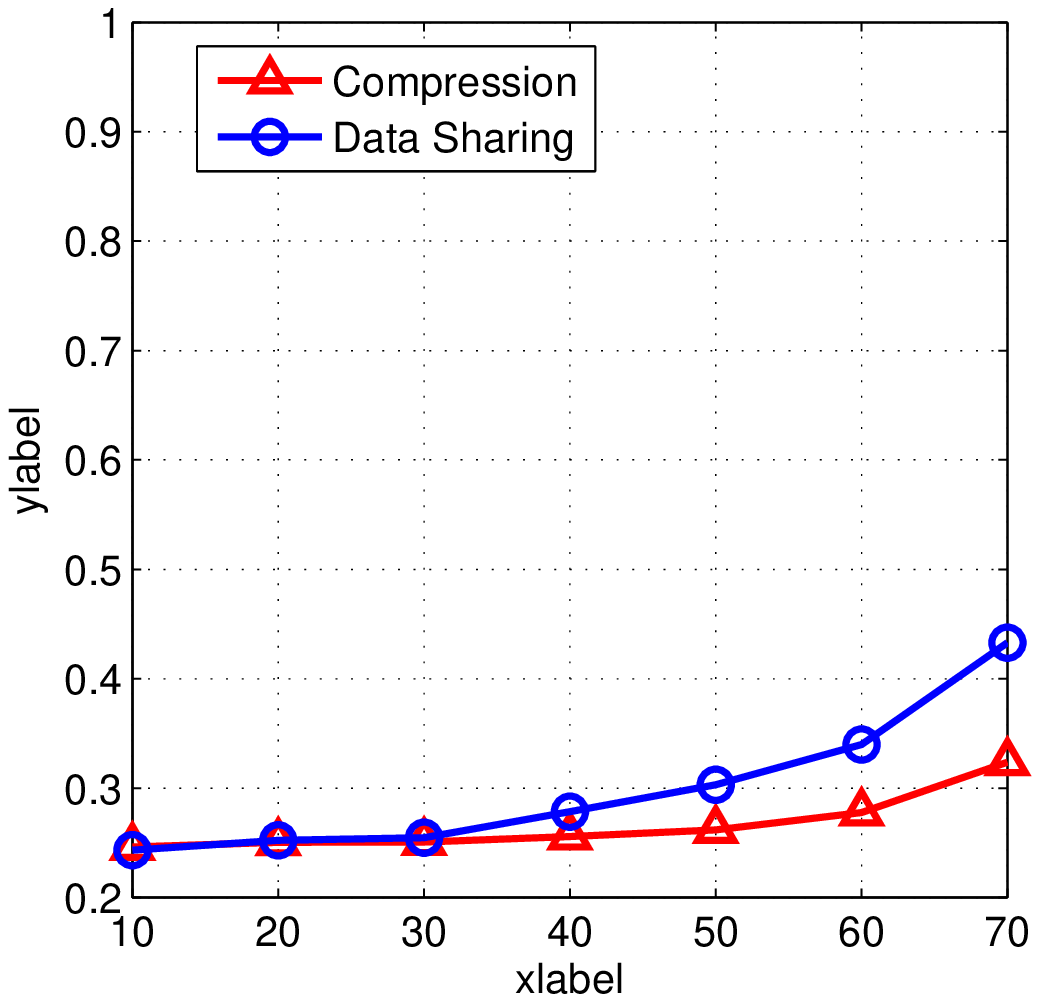}%
\label{sim:bs_u1}}
\hfil
\subfloat[2 users per cell]{\includegraphics[width=0.33\textwidth]{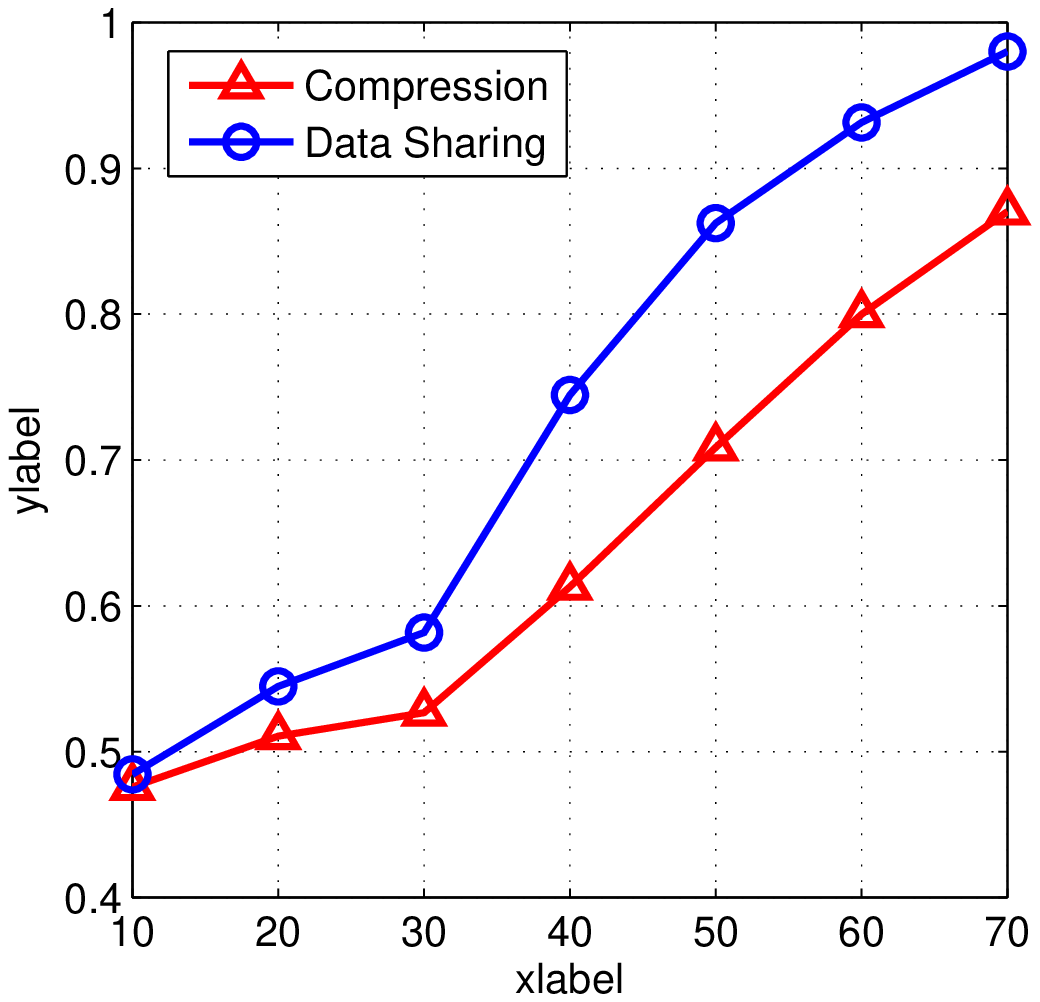}%
\label{sim:bs_u2}}
\hfil
\subfloat[3 users per cell]{\includegraphics[width=0.33\textwidth]{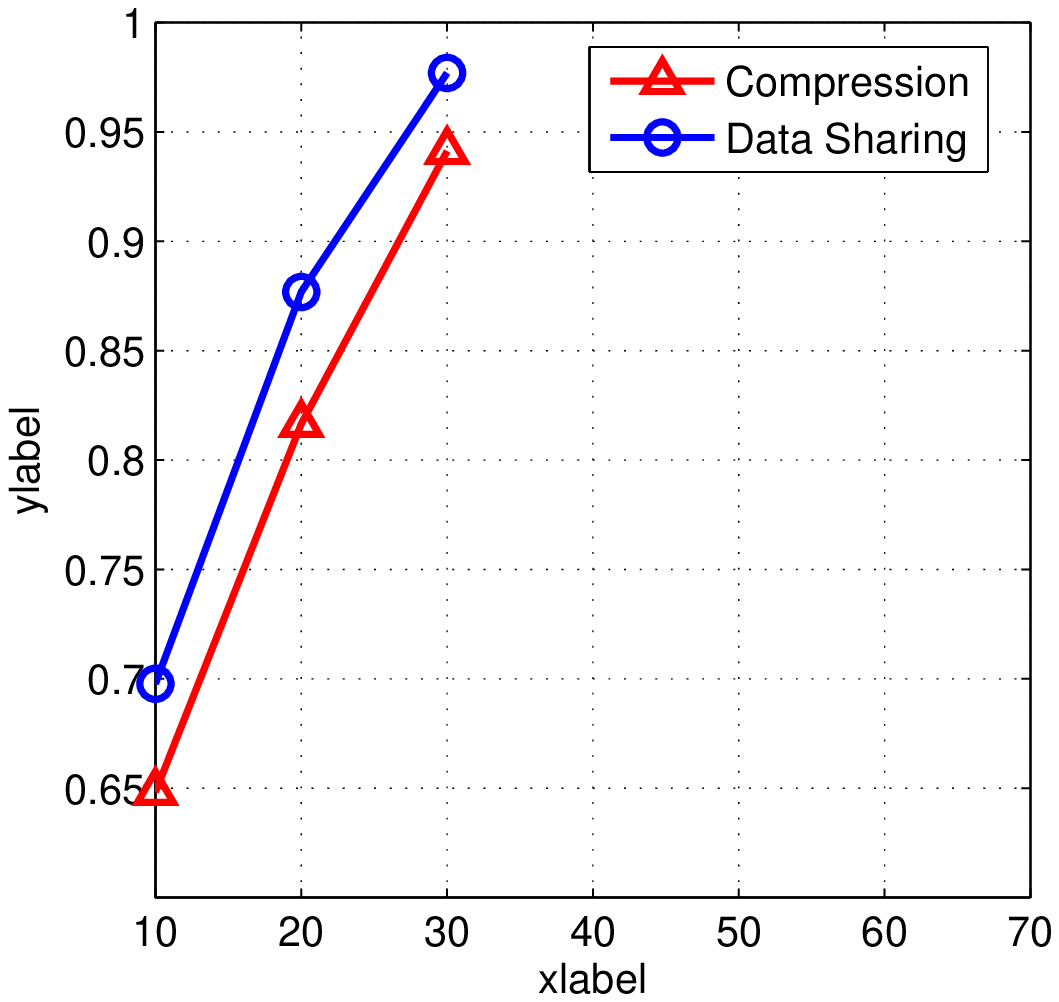}%
\label{sim:bs_u3}}
\caption{Comparison of average fraction of active BSs. Note that the ``Per-Cell CoMP'' scheme considered in Fig.~\ref{sim:total_power} 
corresponds to $100\%$ of active BSs, while the percentage of active BSs for ``Single BS Association'' 
scheme depends on the number of users in each cell, which are $25\%$, $50\%$ and $75\%$ respectively for $1$, $2$ and $3$ users per cell, 
although only those few points corresponding to Fig.~\ref{sim:total_power} are feasible.}
\label{sim:active_bs}
\end{figure*}

In Fig.~\ref{sim:active_bs}, we compare the percentage of active BSs in the data-sharing strategy versus in the compression strategy. 
Similar to Fig.~\ref{sim:total_power}, each point in Fig.~\ref{sim:active_bs} is averaged over $100$ channel realizations. 
As we can see, with higher user target rate and more users to be served, more BSs need to remain active for 
transmission. Also, from Fig.~\ref{sim:active_bs}, it is observed that the compression strategy tends to turn off more BSs than the 
data-sharing strategy.

%
% Note that often IEEE papers with subfigures do not employ subfigure
% captions (using the optional argument to \subfloat), but instead will
% reference/describe all of them (a), (b), etc., within the main caption.

% An example of a floating table. Note that, for IEEE style tables, the 
% \caption command should come BEFORE the table. Table text will default to
% \footnotesize as IEEE normally uses this smaller font for tables.
% The \label must come after \caption as always.
%
%\begin{table}[!t]
%% increase table row spacing, adjust to taste
%\renewcommand{\arraystretch}{1.3}
% if using array.sty, it might be a good idea to tweak the value of
% \extrarowheight as needed to properly center the text within the cells
%\caption{An Example of a Table}
%\label{table_example}
%\centering
%% Some packages, such as MDW tools, offer better commands for making tables
%% than the plain LaTeX2e tabular which is used here.
%\begin{tabular}{|c||c|}
%\hline
%One & Two\\
%\hline
%Three & Four\\
%\hline
%\end{tabular}
%\end{table}

% Note that IEEE does not put floats in the very first column - or typically
% anywhere on the first page for that matter. Also, in-text middle ("here")
% positioning is not used. Most IEEE journals/conferences use top floats
% exclusively. Note that, LaTeX2e, unlike IEEE journals/conferences, places
% footnotes above bottom floats. This can be corrected via the \fnbelowfloat
% command of the stfloats package.

\section{Conclusion}\label{sec:conclusion}

This paper compares the energy efficiency between the data-sharing strategy and the compression strategy in downlink C-RAN. 
We formulate the problem as that of minimizing the total network power consumption subject to user target rate constraints, 
with both the BS power consumption and the backhaul power consumption taken into account. 
By taking advantage of the $\ell_1$-norm reweighting technique and successive convex approximation technique, we 
transform the nonconvex optimization problems into convex form and devise efficient algorithms with provable 
convergence guarantees. 

The main conclusions of this paper are that C-RAN significantly 
improves the range of feasible user data rates in a wireless cellular network, and that 
both data-sharing and compression strategies bring 
much improved energy efficiency to downlink C-RAN as compared to non-optimized CoMP. 
Moreover, between the data-sharing strategy and the compression strategy, either may be preferred 
depending on the different target rate regimes: at low user target rate, data-sharing consumes less power, while at high user target 
rate compression is preferred since the backhaul rate for data-sharing increases significantly as user 
rate increases.

%\balance


\appendices

\section{Proof for Theorem~\ref{thm:1}}\label{apdx:a}

The idea is to show that Algorithm~\ref{alg:data_sharing} 
converges to the stationary point solution of the following problem: 
\begin{align} \label{prob:data_sharing_approx_lim}
\displaystyle \min_{\left\{w_{lk} \right\}} & \quad \sum_{l\in\mathcal{L}} 
\Bigg( \eta_l \sum_{k \in \mathcal{K}} \left\vert w_{lk} \right\vert^{2}  + 
P_{l, \Delta}  \frac{\ln \left( 1 +  \tau_1^{-1} \sum_{k \in \mathcal{K}} \left\vert w_{lk} \right\vert^{2}\right)}{\ln \left( 1 + \tau_1^{-1} \right)}  \nonumber \\ 
&  \hspace{1.4cm} + 
\rho_l \sum_{k\in \mathcal{K}}    \frac{\ln \left( 1 + \tau_2^{-1} \left\vert w_{lk} \right\vert^{2} \right)}{\ln \left( 1 + \tau_2^{-1} \right)}     r_k \Bigg) \\
 \st  & \quad \eqref{sinr_const},  ~~ \eqref{Per_BS_Power} \nonumber 
\end{align}
%in the sense that $\lim_{n \to \infty} d\left( \mathbf{x}^{(n)}, \mathcal{S}\right) = 0$, 
%where $\mathcal{S}$ is the set of KKT points of \eqref{prob:data_sharing_approx_lim} and 
%$d\left( \mathbf{x}^n, \mathcal{S}\right) \triangleq \inf_{\mathbf{s} \in \mathcal{S}} \left\| \mathbf{x}^n - \mathbf{s} \right\|_2$.

First, we note that problem \eqref{prob:data_sharing_approx_lim} differs from the original problem \eqref{prob:data_sharing} in that 
the $\ell_0$-norms in the objective function \eqref{obj_data_sharing} are approximated by the logarithmic functions in 
\eqref{prob:data_sharing_approx_lim}. This approximation stems from the relation that with $x \geq 0$, 
\begin{equation}\label{lim_approx}
\mathbbm{1}_{\left\{x\right\}} = \left\Vert x \right\Vert_0 = \lim_{\tau \to 0} \frac{\ln \left( 1 + x \tau^{-1}\right)}{\ln \left( 1 + \tau^{-1} \right)} ~.
\end{equation}

Now, due to the concavity of $\ln x$, the inequality $\ln x \leq \ln x_0 + x_0^{-1} x - 1$ 
holds for any $x>0, x_0>0$ and achieves equality if and only if $x = x_0$. Hence, we have 
\begin{align} \label{data_ineq}
\hspace{-4mm} \eqref{prob:data_sharing_approx_lim}  \leq &  \sum_{l\in\mathcal{L}} 
\Bigg( \eta_l \sum_{k \in \mathcal{K}} \left\vert w_{lk} \right\vert^{2}   \nonumber \\ 
& \hspace{2mm} + P_{l, \Delta}  \frac{ \ln x_{l} + x_{l}^{-1} \left( 1 +  \tau_1^{-1} \sum_{k \in \mathcal{K}} \left\vert w_{lk} \right\vert^{2} \right) - 1   }{\ln \left( 1 + \tau_1^{-1} \right)}   \nonumber \\
& \hspace{2mm} + \rho_l \sum_{k\in \mathcal{K}}    \frac{ \ln y_{lk} + y_{lk}^{-1} \left(   1 + \tau_2^{-1} \left\vert w_{lk} \right\vert^{2}  \right) - 1 }{\ln \left( 1 + \tau_2^{-1} \right)}     r_k \Bigg) ~ , \nonumber \\
& \hspace{4.2cm} \forall x_{l} >0, \quad y_{lk} > 0 
\end{align}
with equality if and only if 
\begin{align}\label{x_update}
x_{l} = 1 +  \tau_1^{-1} \sum_{k \in \mathcal{K}} \left\vert w_{lk} \right\vert^{2}, \quad
y_{lk} = 1 + \tau_2^{-1} \left\vert w_{lk} \right\vert^{2}. 
\end{align}

Although problem \eqref{prob:data_sharing_approx_lim} is nonconvex due to the logarithmic objective function, 
the function on the right-hand side of \eqref{data_ineq} is a convex quadratic function in $\left\{ w_{lk} \right\}$. 
Based on this fact, we can develop an MM algorithm to solve problem 
\eqref{prob:data_sharing_approx_lim} by solving a sequence of convex optimization problems with the objective function in 
\eqref{prob:data_sharing_approx_lim} replaced by 
the convex function in \eqref{data_ineq} and iteratively updating the parameters $x_{l}, y_{lk}$ 
according to \eqref{x_update}.  
Comparing such MM algorithm with Algorithm~\ref{alg:data_sharing}, it is easy to see that 
Algorithm~\ref{alg:data_sharing} reduces to the MM algorithm 
if the reweighting function in \eqref{wgt_update_data} is chosen as \eqref{eq:reweight}. 

It is known in the literature \cite{Meisam13} that an MM algorithm is guaranteed to converge to the stationary point solutions of the original problem if 
the approximate function satisfies the following conditions: 1) it is continuous, 
2) it is a tight upper bound of the original objective function and 
3) it has the same first-order derivative as the original objective function at the point where the upper bound is tight. 
It is easy to check that the function in \eqref{data_ineq} satisfies all these sufficient conditions. 
Thus, Algorithm~\ref{alg:data_sharing}, which is equivalent to an MM algorithm, must converge.

\section{Proof for Theorem~\ref{thm:2}}\label{apdx:b}

Similar to the proof for Theorem~\ref{thm:1}, the idea is show that Algorithm~\ref{alg:compression} converges to the stationary point 
solution of the following optimization problem: 
\begin{align} \label{prob:compression_approx}
\displaystyle  \mini_{\left\{w_{lk}, q_l\right\}} & \quad \sum_{l\in\mathcal{L}} 
\Bigg( \eta_l \left( \sum_{k \in \mathcal{K}} \left\vert w_{lk} \right\vert^{2} + q_l^2 \right) \nonumber \\ 
& \hspace{6mm}+ 
 P_{l, \Delta} \frac{\ln\left( 1 + \tau_3^{-1} \left( \sum_{k \in \mathcal{K}} \left\vert w_{lk} \right\vert^{2} + q_l^2  \right) \right)}{\ln\left( 1 + \tau_3^{-1}  \right)} \nonumber \\
 & \hspace{6mm} + \rho_l \log_2 \left( 1 + \frac{\Gamma_q \sum_{k \in \mathcal{K}} \left\vert w_{lk} \right\vert^{2}}{q_l^2}  \right) \Bigg) \\
 \sbto  &  \quad \eqref{sinr_const_compression}, ~~ \eqref{compression_power_const} ~,\nonumber
\end{align}
which is a logarithmic approximation to the original problem \eqref{prob:compression}. 

Problem \eqref{prob:compression_approx} is nonconvex due to the logarithmic functions in its objective function. 
However, we can develop an MM algorithm to solve \eqref{prob:compression_approx} by solving a sequence of 
convex optimization problems with the objective function in \eqref{prob:compression_approx} replaced by its upper bound shown below 
\begin{align}\label{func:major_comp}
%& \hspace{-3mm}\eqref{prob:compression_approx} \leq \nonumber \\ 
&  \sum_{l\in\mathcal{L}} 
\Bigg( \eta_l \left( \sum_{k \in \mathcal{K}} \left\vert w_{lk} \right\vert^{2} + q_l^2 \right)  - 2 \rho_l \log_2 q_l \nonumber \\ 
& \quad + 
 P_{l, \Delta} \frac{\ln z_{l} + z_{l}^{-1} \left( 1 +  \tau_3^{-1}\left( \sum_{k \in \mathcal{K}} \left\vert w_{lk} \right\vert^{2} + q_l^2\right)\right) - 1}{\ln\left( 1 + \tau_3^{-1}  \right)} \nonumber \\
 & \quad + 
\rho_l \left(\log_2 \lambda_l  +  \frac{ q_l^2 + \Gamma_q \sum_{k \in \mathcal{K}} \left\vert w_{lk} \right\vert^{2}}{\lambda_l \ln 2} - 
\frac{1}{\ln 2} \right) %\nonumber \\ 
  \Bigg)
\end{align}
and iteratively updating the parameters $z_{l}$ as $z_{l} = 1 +  \tau_3^{-1}\left( \sum_{k \in \mathcal{K}} \left\vert w_{lk} \right\vert^{2} + q_l^2 \right)$ 
and $\lambda_l$ as \eqref{lambda_update}.  
It is easy to see that such MM algorithm is equivalent to Algorithm~\ref{alg:compression} with the reweighting 
function in \eqref{wgt_mu} chosen as \eqref{eq:reweight}. 
We can also easily verify that the majorizing function in the right-hand side of \eqref{func:major_comp} 
satisfies all the sufficient conditions in \cite{Meisam13} for the convergence guarantee of MM algorithm. 
Hence, Algorithm~\ref{alg:compression} must converge.

%\section*{Acknowledgment}



%\bibliographystyle{IEEEtran}
%\begin{thebibliography}{1}
  %0.5em minus 0.4em\relax Harlow, England: Addison-Wesley, 1999.
%
%\end{thebibliography}

\bibliographystyle{IEEEtran}
%\bibliography{strings,refs}

\bibliography{IEEEabrv,myref}

%
%\newpage
%\include{response/response}

%
%

\begin{IEEEbiography}[{\includegraphics[width=1in,height=1.25in,clip,keepaspectratio]{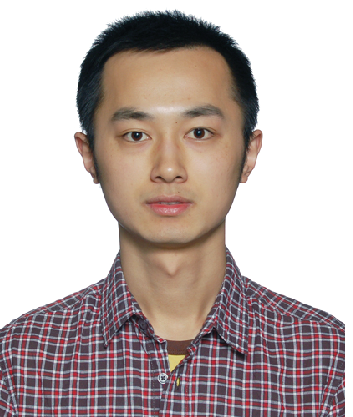}}]
{Binbin Dai} (S'12) received the B.E. degree in Information Science and Engineering from Chien-Shiung Wu Honors College, Southeast University, Nanjing, China, in 2011 and M.A.Sc degree in Electrical and Computer Engineering from University of Toronto, Toronto, Ontario, Canada, in 2014. He is currently working towards the Ph.D degree with the Department of Electrical and Computer Engineering, University of Toronto, Toronto, Ontario, Canada. His research interests include optimization, wireless communications and signal processing. 
\end{IEEEbiography}

\begin{IEEEbiography}[{\includegraphics[width=1in,height=1.25in,clip,keepaspectratio]{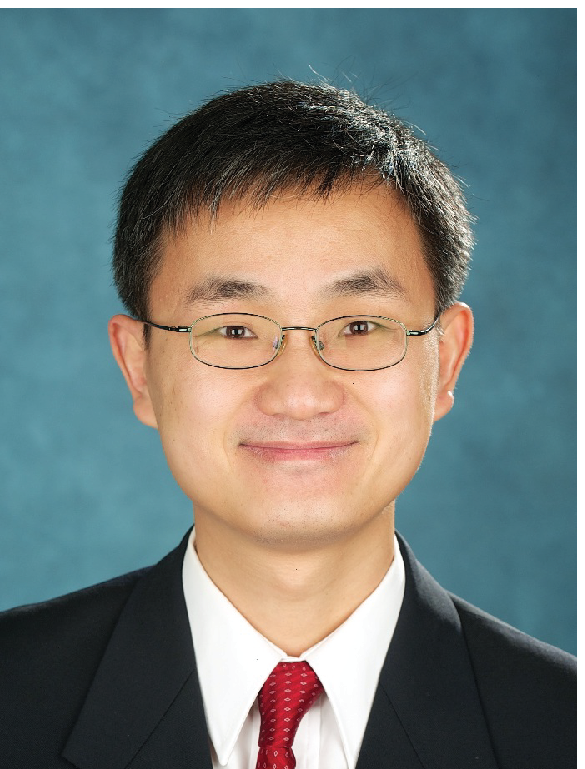}}]
{Wei Yu} (S'97-M'02-SM'08-F'14) received the B.A.Sc. degree in Computer Engineering and Mathematics from the University of Waterloo, Waterloo, Ontario, Canada in 1997 and M.S. and Ph.D. degrees in Electrical Engineering from Stanford University, Stanford, CA, in 1998 and 2002, respectively. Since 2002, he has been with the Electrical and Computer Engineering Department at the University of Toronto, Toronto, Ontario, Canada, where he is now Professor and holds a Canada Research Chair (Tier 1) in Information Theory and Wireless Communications. His main research interests include information theory, optimization, wireless communications and broadband access networks.

Prof. Wei Yu currently serves on the IEEE Information Theory Society Board of Governors (2015-17). He is an IEEE Communications Society Distinguished Lecturer (2015-16). He served as an Associate Editor for \textsc{IEEE Transactions on Information Theory} (2010-2013), as an Editor for \textsc{IEEE Transactions on Communications} (2009-2011), as an Editor for \textsc{IEEE Transactions on Wireless Communications} (2004-2007), and as a Guest Editor for a number of special issues for the \textsc{IEEE Journal on Selected Areas in Communications} and the \textsc{EURASIP Journal on Applied Signal Processing}. He was a Technical Program co-chair of the IEEE Communication Theory Workshop in 2014, and a Technical Program Committee co-chair of the Communication Theory Symposium at the IEEE International Conference on Communications (ICC) in 2012. He was a member of the Signal Processing for Communications and Networking Technical Committee of the IEEE Signal Processing Society (2008-2013). Prof. Wei Yu received a Steacie Memorial Fellowship in 2015, an IEEE Communications Society Best Tutorial Paper Award in 2015, an IEEE ICC Best Paper Award in 2013, an IEEE Signal Processing Society Best Paper Award in 2008, the McCharles Prize for Early Career Research Distinction in 2008, the Early Career Teaching Award from the Faculty of Applied Science and Engineering, University of Toronto in 2007, and an Early Researcher Award from Ontario in 2006. He was named a Highly Cited Researcher by Thomson Reuters in 2014.

%Prof. Wei Yu is a Fellow of IEEE. He is a registered Professional Engineer in Ontario.
\end{IEEEbiography}

\end{document}